\documentclass[sigconf, screen]{acmart}

\usepackage{algorithmic}
\usepackage{graphicx}
\usepackage{textcomp}
\usepackage{xcolor}
\usepackage{fancyhdr}
\usepackage[normalem]{ulem}
\usepackage{xspace}
\usepackage{scalerel} 
\usepackage{hhline}
\usepackage{upgreek}
\usepackage{footnote}
\usepackage{enumitem}
\usepackage{graphicx}
\usepackage[bottom]{footmisc}
\usepackage{hyperref}
\usepackage{url}
\usepackage{textcomp}
\usepackage{array}
\usepackage{multicol}
\usepackage{multirow}
\usepackage{courier}
\usepackage{tikz} 
\usepackage{dblfloatfix} 
\usepackage{longfbox}
\usepackage{balance}
\usepackage[ddmmyyyy]{datetime}
\usepackage{breakcites}
\usepackage[colorinlistoftodos]{todonotes}
\definecolor{dollarbill}{rgb}{0.52, 0.73, 0.4}

\newcommand\affileth{$^1$}
\newcommand\affilsnu{$^2$}
\newcommand\affilknu{$^3$}

\newcommand{\sect}[1]{{Section~#1}\xspace} % for referencing section
\newcommand{\head}[1]{{\noindent\textbf{#1.}\xspace}} % for heading of a paragraph
\newcommand{\figs}[1]{{Figures~#1}\xspace} % for referencing figures
\newcommand{\fig}[1]{{Figure~#1}\xspace} % for referencing figure
\newcommand{\tab}[1]{{Table~#1}\xspace} % for referencing table
\newcolumntype{?}{!{\vrule width 1pt}} % for columns
\newcolumntype{;}{!{\vrule width 0.5pt}}
\newcolumntype{P}[1]{>{\centering\arraybackslash}p{#1}}
\DeclareRobustCommand\bcirc[1]{\tikz[baseline=(char.base)]{
           \node[shape=circle,draw,inner sep=0pt,fill=black, text=white] (char) {#1};}}
\newcommand{\prr}{{PR$^{2}$}\xspace}
\newcommand{\arr}{{AR$^{2}$}\xspace}
\newcommand{\base}{\textsf{Baseline}\xspace}
\newcommand{\prssd}{\textsf{PR$^{2}$}\xspace}
\newcommand{\arssd}{\textsf{AR$^{2}$}\xspace}
\newcommand{\parssd}{\textsf{PnAR$^{2}$}\xspace}
\newcommand{\norr}{\textsf{NoRR}\xspace}
\newcommand{\cread}{\texttt{CACHE}~\texttt{READ}\xspace}
\newcommand{\pread}{\texttt{PAGE}~\texttt{READ}\xspace}

\newcommand\stg{\textsf{stg\_0}\xspace}
\newcommand\hm{\textsf{hm\_0}\xspace}
\newcommand\prn{\textsf{prn\_1}\xspace}
\newcommand\proj{\textsf{proj\_1}\xspace}
\newcommand\mds{\textsf{mds\_1}\xspace}
\newcommand\usr{\textsf{usr\_1}\xspace}
\newcommand\yc[1]{\textsf{YCSB-#1}\xspace}

\newcommand{\vth}{$\text{V}_\text{TH}$\xspace} % for threshold voltage
\newcommand{\vref}[1]{$\text{V}_{\text{REF}#1}$\xspace} % for read-reference voltage
\newcommand{\vrr}[1]{$\text{V}_{\text{RR}#1}$\xspace}
\newcommand{\vopt}{$\text{V}_{\text{OPT}}$\xspace}
\newcommand{\usec}{$\mu$s\xspace} % for micro second
\newcommand{\tr}{\texttt{tR}\xspace} % for chip-level read latency
\newcommand{\tprog}{\texttt{tPROG}\xspace} % for chip-level write latency
\newcommand{\tbers}{\texttt{tBERS}\xspace} % for erase latency
\newcommand{\tpre}{\texttt{tPRE}\xspace} % for precharge latency
\newcommand{\teval}{\texttt{tEVAL}\xspace} % for evaluation latency
\newcommand{\tdisch}{\texttt{tDISCH}\xspace} % for discharge latency
\newcommand{\tread}{\texttt{tREAD}\xspace} % for ssd-level read latency
\newcommand{\trr}{\texttt{tRETRY}\xspace} % for read-retry latency
\newcommand{\nrr}{$N_\text{RR}$\xspace}
\newcommand{\nerr}{$M_\text{ERR}$\xspace}
\newcommand{\derr}{$\Delta{M}_\text{ERR}$\xspace}
\newcommand{\nerri}[2]{$M_\text{ERR}$(#1,~#2)\xspace}
\newcommand{\derri}[2]{$\Delta{M}_\text{ERR}$(#1,~#2)\xspace}
\newcommand{\pec}{$PEC$\xspace}
\newcommand{\tret}{$t_\text{RET}$\xspace}
\newcommand{\tecc}{\texttt{tECC}\xspace} % for discharge latency
\newcommand{\tdma}{\texttt{tDMA}\xspace} % for dma latency
\newcommand{\degreec}[1]{~\ensuremath{#1^\circ}C\xspace} % for degree celsius

\AtBeginDocument{%
  \providecommand\BibTeX{{%
    \normalfont B\kern-0.5em{\scshape i\kern-0.25em b}\kern-0.8em\TeX}}}

\setcopyright{acmlicensed}
\acmPrice{15.00}
\acmDOI{10.1145/3445814.3446719}
\acmYear{2021}
\copyrightyear{2021}
\acmSubmissionID{asplos21main-p278-p}
\acmISBN{978-1-4503-8317-2/21/04}
\acmConference[ASPLOS '21]{Proceedings of the 26th ACM International Conference on Architectural Support for Programming Languages and Operating Systems}{April 19--23, 2021}{Virtual, USA}
\acmBooktitle{Proceedings of the 26th ACM International Conference on Architectural Support for Programming Languages and Operating Systems (ASPLOS '21), April 19--23, 2021, Virtual, USA}

\begin{document}

%argument: (short) title to be printed on the header
\title[Reducing Solid-State Drive Read Latency by Optimizing Read-Retry]{Reducing Solid-State Drive Read Latency\\by Optimizing Read-Retry}

%ASPLOS asks for individual author entries

\author{Jisung Park\affileth \quad Myungsuk Kim$^{2,3}$ \quad Myoungjun Chun\affilsnu \quad Lois Orosa\affileth \quad Jihong Kim\affilsnu \quad Onur Mutlu\affileth}
\affiliation{
\institution{\vspace{1em}}
\institution{\affileth{}ETH Z\"urich\quad\quad
\affilsnu{}Seoul National University\quad\quad \affilknu{}Kyungpook National University}
\country{Switzerland \quad\quad\quad\quad Republic of Korea \quad{  }\quad\quad\quad\quad{    }\quad Republic of Korea \quad\quad}
\institution{\vspace{1em}}
}

% \author{Jisung Park}
% \email{jisung.park@safari.ethz.ch}
% \affiliation{%
%   \institution{ETH Z\"urich}
%   \country{Switzerland}}

% \author{Myungsuk Kim}
% \email{morssola75@gmail.com}
% \affiliation{%
%   \institution{Kyungpook National University}
%   \country{Republic of Korea}}

% \author{Myoungjun Chun}
% \email{mjchun@davinci.snu.ac.kr}
% \affiliation{%
%   \institution{Seoul National University}
%   \country{Republic of Korea}}
  
% \author{Lois Orosa}
% \email{lois.orosa.nogueira@gmail.com}
% \affiliation{%
%   \institution{ETH Z\"urich}
%   \country{Switzerland}}

% \author{Jihong Kim}
% \email{jihong@davinci.snu.ac.kr}
% \affiliation{%
%   \institution{Seoul National University}
%   \country{Republic of Korea}}

% \author{Onur Mutlu}
% \email{omutlu@gmail.com}
% \affiliation{%
%   \institution{ETH Z\"urich}
%   \country{Switzerland}}

%If a single-line author list is not possible, 1) first try to initial each author's given name and 2) just list the first author only.
%\renewcommand{\shortauthors}{Park, et al.}
\renewcommand{\shortauthors}{Jisung Park, Myungsuk Kim, Myoungjun Chun, Lois Orosa, Jihong Kim, and Onur Mutlu}
\renewcommand{\authors}{Jisung Park, Myungsuk Kim, Myoungjun Chun, Lois Orosa, Jihong Kim, and Onur Mutlu}

\begin{abstract}
3D NAND flash memory with advanced multi-level cell techniques provides high storage density, but suffers from significant performance degradation due to a large number of read-retry operations.
Although the read-retry mechanism is essential to ensuring the reliability of modern NAND flash memory, it can significantly increase the read latency of an SSD by introducing multiple retry steps that read the target page again with adjusted read-reference voltage values. 
Through a detailed analysis of the read mechanism and rigorous characterization of 160 real 3D NAND flash memory chips, we find new opportunities to reduce the read-retry latency by exploiting two advanced features widely adopted in modern NAND flash-based SSDs: 1) the \cread command and 2) strong ECC engine. 
First, we can reduce the read-retry latency using the advanced \cread command that allows a NAND flash chip to perform consecutive reads in a pipelined manner. 
Second, there exists a large ECC-capability margin in the final retry step that can be used for reducing the chip-level read latency. 
Based on our new findings, we develop two new techniques that effectively reduce the read-retry latency: 1) \emph{\underline{P}ipelined \underline{R}ead-\underline{R}etry (\prr)} and 2) \emph{\underline{A}daptive \underline{R}ead-\underline{R}etry (\arr)}. 
\prr reduces the latency of a read-retry operation by pipelining consecutive retry steps using the \cread command. 
\arr shortens the latency of each retry step by dynamically reducing the chip-level read latency depending on the current operating conditions that determine the ECC-capability margin. 
Our evaluation using twelve real-world workloads shows that our proposal improves SSD response time by up to 31.5\% (17\% on average) over a state-of-the-art baseline with only small changes to the SSD controller.
\end{abstract}

% The code below is generated by the tool at http://dl.acm.org/ccs.cfm.
\vspace{10pt}
\begin{CCSXML}
<ccs2012>
<concept>
<concept_id>10010583.10010588.10010592</concept_id>
<concept_desc>Hardware~External storage</concept_desc>
<concept_significance>500</concept_significance>
</concept>
</ccs2012>
\end{CCSXML}


% Keywords. Separate the keywords with commas.

% Print page numbers
\settopmatter{printfolios=true}

\maketitle

% Remove the page number from the first page
\thispagestyle{empty}

%Main sections
\section{Introduction\label{sec:intro}}
NAND flash memory is the prevalent technology for architecting storage devices in modern computing systems to meet high storage-capacity and I/O-performance requirements.
3D NAND technology and advanced multi-level cell (MLC) techniques enable continuous increase of storage density, but they negatively affect the reliability of modern NAND flash chips. 
NAND flash memory stores data as the \emph{threshold voltage (\vth) level} of each flash cell, which depends on the amount of charge in the cell. 
New cell designs and organizations in 3D NAND flash memory cause a flash cell to leak its charge more easily~\cite{cai-insidessd-2018, luo-hpca-2018, luo-sigmetrics-2018}.
In addition, MLC technology significantly reduces the margin between different \vth levels used to store multiple bits in a single flash cell.
Consequently, the \vth level of a 3D NAND flash cell with advanced MLC techniques (e.g., triple-level cell (TLC)~\cite{kang-isscc-2016, cai-procieee-2017} or quad-level cell (QLC)~\cite{huh-iccss-2020, kim-isscc-2020}) can quickly shift beyond the read-reference voltage \vref{} (i.e., the voltage used to distinguish between cell \vth levels) after programming, which results in an error when reading the cell.

To guarantee the reliability of stored data, a modern SSD commonly adopts two main approaches.
First, a modern SSD employs strong \emph{error-correcting codes (ECC)} that can detect and correct several tens of raw bit errors (e.g., 72 bits per 1-KiB codeword~\cite{micron-flyer-2016}).
Second, when ECC fails to correct all bit errors, the SSD controller performs a \emph{read-retry operation} that reads the erroneous page\footnote{A NAND flash memory concurrently reads and writes multiple cells at a page (e.g., 16 KiB) granularity (see \sect{\ref{subsec:nand_opr}}).} again with \emph{slightly-adjusted} \vref{} values. 
Since bit errors occur when the \vth levels of flash cells shift beyond the \vref{} values, sensing the cells with appropriately-shifted \vref{} values can greatly reduce the number of raw bit errors~\cite{cai-procieee-2017, cai-insidessd-2018, cai-hpca-2017, cai-date-2013, cai-dsn-2015, cai-hpca-2015, cai-iccd-2013, cai-inteltechj-2013, cai-sigmetrics-2014, luo-ieeejsac-2016, luo-hpca-2018, luo-sigmetrics-2018, shim-micro-2019}. 

Although read-retry is essential to ensuring the reliability of modern NAND flash memory, it comes at the cost of significant performance degradation.
A read-retry operation \emph{repeats} a retry step until it finds \vref{} values that allow the page's raw bit-error rate (RBER) to be lower than the ECC correction capability (i.e., the number of errors correctable) or finds for sure that the page cannot be read without errors.
Recent work~\cite{shim-micro-2019} shows that a modern SSD with long \emph{retention age values} (i.e., how long data is stored after it is programmed) and high program/erase (P/E) cycles (i.e., how many writes/erases are performed) suffers from a large number of read-retry operations that increase the read latency linearly with the number of retry steps.
Our experimental characterization using 160 real 3D TLC NAND flash chips, in this work, shows that a read frequently incurs \emph{multiple} retry steps even under modest operating conditions.
For example, under a 3-month data retention age at \emph{zero} P/E cycles (i.e., at the beginning of SSD lifetime), we observe that every read requires more than three retry steps.

Prior works~\cite{cai-hpca-2015, cai-iccd-2013, luo-ieeejsac-2016, luo-hpca-2018, luo-sigmetrics-2018, nie-dac-2020, shim-micro-2019} attempt to reduce the number of retry steps by quickly identifying near-optimal \vref{} values, but read-retry is a fundamental problem that is \emph{difficult to completely eliminate} in modern SSDs.
For example, an existing technique~\cite{shim-micro-2019} reads a page using \vref{} values that have been recently used for a read-retry operation on other pages exhibiting similar error characteristics with the page to read.
Doing so significantly reduces the number of retry steps by starting a read (and retry) operation with the \vref{} values close to the \emph{optimal} read-reference voltage (\vopt) values.
However, this technique cannot completely avoid read-retry: \emph{every read} still incurs at least three retry steps in an aged SSD~\cite{shim-micro-2019}.
This is because, in modern NAND flash memory, the \vth levels of flash cells change quickly and significantly over time, which makes it extremely difficult to identify the exact \vref{} values that can avoid read-retry before reading the target page.

In this paper, we identify new opportunities to reduce the read-retry latency by exploiting two advanced architectural features widely adopted in modern SSDs: 1) the \emph{\cread command}~\cite{leong-uspatent-2008, macronix-technote-2013, micron-technote-2004} and 2) \emph{strong ECC engine}~\cite{cai-procieee-2017, cai-insidessd-2018}.
First, we find that it is possible to reduce the total execution time of a read-retry operation using the \cread command that allows a NAND flash chip to perform consecutive reads in a pipelined manner. 
Since each retry step is effectively the same as a regular page read, the \cread command also enables concurrent execution of consecutive retry steps in a read-retry operation.

Second, we find that a large ECC-capability margin exists in the final retry step.
Although a read-retry occurs when the read page's RBER exceeds the ECC capability (i.e., when there is no ECC-capability margin), once a read-retry operation succeeds, it allows the page to be eventually read \emph{without} any uncorrectable errors (i.e., there exists a \emph{positive} ECC-capability margin in the final retry step).
We hypothesize that the ECC-capability margin is large due to two reasons.
First, a modern SSD uses \emph{strong} ECC that can correct several tens of raw bit errors in a codeword.
Second, in the final retry step, the page can be read by using \emph{near-optimal} \vref{} values that drastically decrease the page's RBER.
If we can leverage the large ECC-capability margin to reduce the \emph{page-sensing latency \tr}, it allows not only the final retry step to quickly read the page without uncorrectable errors but also the earlier retry steps (which would fail anyway even with the default \tr) to be finished more quickly. 
To validate our hypothesis, we characterize 1) the ECC-capability margin in each retry step and 2) the impact of reducing \tr on the page's RBER, using 160 real 3D TLC NAND flash chips.
The results show that we can safely reduce \tr of each retry step by 25\% even under the worst-case operating conditions prescribed by manufacturers (e.g., a 1-year data retention age~\cite{cox-fms-2018} at 1.5K P/E cycles~\cite{micron-flyer-2016}).

Based on our findings, we develop two new read-retry mechanisms that effectively reduce the read-retry latency. 
First, we propose \emph{\underline{P}ipelined \underline{R}ead \underline{R}etry (\prr)} that performs consecutive retry steps in a pipelined manner using the \cread command.
Unlike the regular read-retry mechanism that starts a retry step \emph{after} finishing the previous retry step, \prr performs page sensing of a retry step during data transfer of the previous retry step, which removes data transfer and ECC decoding from the critical path of a read-retry operation, reducing the latency of a retry step by 28.5\%.
Second, we introduce \emph{\underline{A}daptive \underline{R}ead \underline{R}etry (\arr)} that performs each retry step with reduced page-sensing latency (\tr), leading to a further 25\% latency reduction even under the worst-case operating conditions.
Since reducing \tr inevitably increases the read page's RBER, an excessive \tr reduction can potentially cause the final retry step to fail to read the page without uncorrectable errors. 
This, in turn, introduces one or more additional retry steps, which could increase the overall read latency.
To avoid increasing the number of retry steps, \arr uses the best \tr value for a certain operating condition that we find via extensive and rigorous characterization of 160 real 3D NAND flash chips. 

Our two techniques require only small modifications to the SSD controller or firmware but no change to underlying NAND flash chips.
This makes our techniques easy to integrate into an SSD along with existing techniques that aim to reduce the \emph{number} of retry steps per read-retry operation~\cite{cai-hpca-2015, cai-iccd-2013, luo-ieeejsac-2016, luo-hpca-2018, luo-sigmetrics-2018, nie-dac-2020, shim-micro-2019}.
Our evaluation using twelve real-world workloads shows that our two techniques, when combined, significantly improve the SSD response time by up to 50.8\% (35.2\% on average) in a baseline high-end SSD. 
Compared to a state-of-the-art research baseline~\cite{shim-micro-2019}, our proposal reduces SSD response time by up to 31.5\% (17\% on average) in read-dominant workloads.

This paper makes the following key contributions:
\vspace{-20pt}
\begin{itemize}[leftmargin=*]
    \item To our knowledge, this work is the first to identify new opportunities to reduce the latency of each retry step by exploiting advanced architectural features widely adopted in modern SSDs.
    \item Through extensive and rigorous characterization of 160 real 3D TLC NAND flash chips, we make three new observations on modern NAND flash memory.
    First, a read-retry occurs frequently even under modest operating conditions (\sect{\ref{subsec:rr_behavior}}).
    Second, when a read-retry occurs, there is a large ECC-capability margin in the final retry step even under the worst-case operating conditions (\sect{\ref{subsec:margin}}).
    Third, there is substantial margin in read-timing parameters, which enables safe reduction of the page-sensing latency in a read-retry operation (\sect{\ref{subsec:reduce_rt}}). 
    \item Based on our findings and characterization results, we propose two new techniques, \prr and \arr, which effectively reduce the latency of each retry step, thereby reducing overall read latency. 
    Our techniques require only very small changes to the SSD controller or firmware.
    By reducing the latency of each retry step while keeping the same number of retry steps during a flash read, our proposal effectively complements existing techniques~\cite{cai-hpca-2015, cai-iccd-2013, luo-ieeejsac-2016, luo-hpca-2018, luo-sigmetrics-2018, nie-dac-2020, shim-micro-2019} that aim to reduce the number of retry steps, as we empirically demonstrate (\sect{\ref{sec:eval}}).
\end{itemize}
\pagebreak

\section{Background}\label{sec:bg}
We provide brief background on relevant aspects of NAND flash memory necessary to understand the rest of the paper.

\subsection{\textbf{NAND Flash Organization}\label{subsec:nand_og}}
\begin{figure}[!h]
	\centering
	\vspace{-.5em}
	\includegraphics[width=0.94\linewidth]{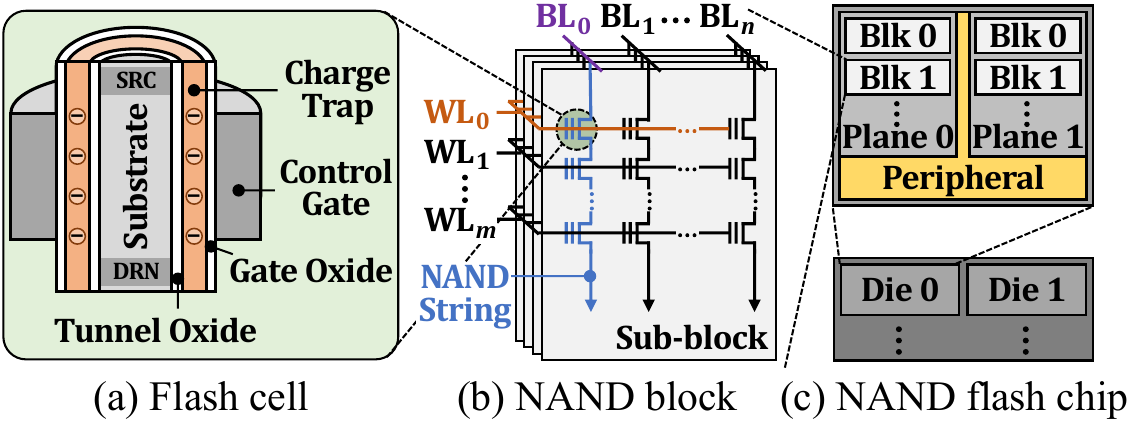}
	\vspace{-1.3em}
	\caption{Organization of 3D NAND flash memory.}
	\vspace{-1em}
	\label{fig:nand_organization}
\end{figure}

NAND flash memory is hierarchically organized.
\fig{\ref{fig:nand_organization}} illustrates the organization of a 3D NAND flash chip.
Multiple (e.g., 24 to 176) flash cells (\fig{\ref{fig:nand_organization}(a))} are vertically stacked and form a \emph{NAND string} (\fig{\ref{fig:nand_organization}(b)}) that is connected to a bitline (BL).
NAND strings at different BLs compose a \emph{sub-block}.
The control gate of each cell at the same vertical location in a sub-block is connected to the same wordline (WL), which makes all the cells at the same WL operate concurrently.
A \emph{block} consists of several (e.g., 4 to 8) sub-blocks, and thousands (e.g., 3,776~\cite{kang-isscc-2016}) of blocks constitute a \emph{plane}.
A NAND flash chip contains multiple \emph{dies} (\fig{\ref{fig:nand_organization}(c)}), each of which comprises multiple planes (e.g., two or four planes per die~\cite{huh-iccss-2020}).
Dies in a NAND flash chip can operate independently of each other, while planes in a die can concurrently operate under limited conditions as they usually share the same row decoder~\cite{hu-ics-2011}.

A flash cell encodes bit data using its threshold voltage (\vth) level. 
As shown in \fig{\ref{fig:nand_organization}(a)}, a flash cell has a special material, called a charge trap\footnote{It is also possible to design 3D NAND flash memory with floating-gate cells~\cite{xiong-acmtos-2018}, but most 3D NAND flash chips adopt cylindrical charge-trap cells (e.g., TCAT~\cite{jang-vlsi-2009}, p-BICs~\cite{katsumata-vlsi-2009}, and SMArT~\cite{choi-iedm-2012})~\cite{cai-procieee-2017, cai-insidessd-2018, shim-micro-2019, luo-sigmetrics-2018, luo-hpca-2018}.}, which can hold electrons without power supply.
The larger the number of electrons in the charge trap, the higher the cell's \vth level.
In single-level cell (SLC) NAND flash memory, for example, a cell can encode one bit of data by encoding its high and low \vth levels as `0' and `1', respectively.

\subsection{NAND Flash Operation\label{subsec:nand_opr}}
Three basic operations enable access to NAND flash memory: 1)~program, 2) erase, and 3) read.

\head{Program and Erase Operations}
A program operation \emph{injects} electrons into a cell's charge trap from the substrate by applying a high voltage ($>$ 20 V) to the WL, which increases the cell's \vth level (i.e., program operation can only change a cell's data from `1' to `0' assuming the SLC encoding described above). 
As a set of flash cells are connected to a single WL in NAND flash memory (i.e., the same voltage is applied to the control gate of every cell in the same WL), data is written at \emph{page granularity} (e.g., 16~KiB) such that each cell at the same WL stores one bit of the page.

An erase operation \emph{ejects} electrons from a cell's charge trap by applying a high voltage ($>$ 20 V) to the substrate, which decreases the cell's \vth level.
As program and erase operations are \emph{unidirectional}, a page needs to be erased first to program data (\emph{erase-before-write}).
A NAND flash chip performs an erase operation at block granularity (for cost reasons).
This leads to a high erase bandwidth because a block consists of hundreds (e.g., 576~\cite{kang-isscc-2016}) or thousands (e.g., 1,472~\cite{kim-isscc-2020}) of pages, but also causes the erase
latency \tbers to be much longer than program latency \tprog (e.g., 3.5~ms vs. 660~\usec~\cite{kang-isscc-2016}).

\begin{figure}[h]
    \centering
    \vspace{-.5em}
    \includegraphics[width=1\linewidth]{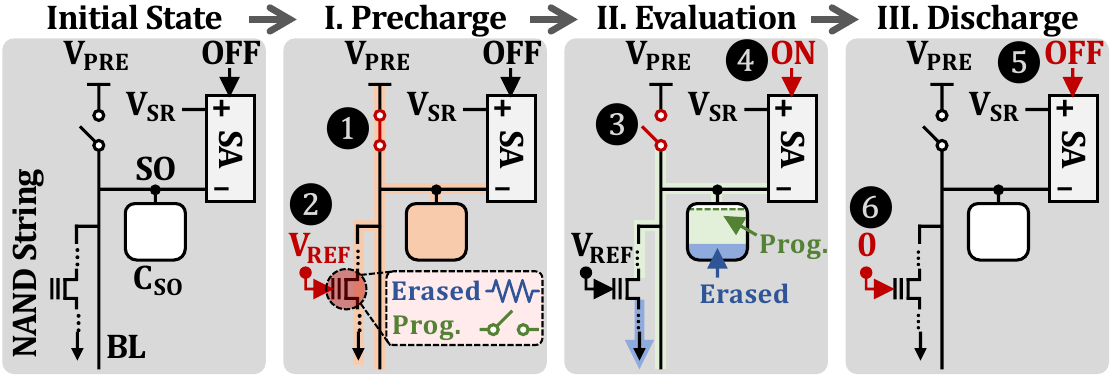}
    \vspace{-2.3em}
    \caption{Read mechanism of NAND flash memory.}
    \vspace{-1em}
    \label{fig:sensing}
\end{figure}

\head{Read Operation}
NAND flash memory determines a cell's data (i.e., the cell's \vth level) by identifying whether current flows through the corresponding BL.
\fig{\ref{fig:sensing}} depicts the read mechanism of NAND flash memory that consists of three phases: 
1) \emph{precharge}, 2) \emph{evaluation}, and 3) \emph{discharge}~\cite{micheloni-2010}.

In the precharge phase (\textbf{I} in \fig{\ref{fig:sensing}}), a NAND flash chip charges each target BL and its sense-out (SO) capacitor $\text{C}_\text{SO}$ to a specific voltage $\text{V}_\text{PRE}$ (\bcirc{1}).
The chip also applies the read-reference voltage \vref{} to the target cell (i.e., WL) at the same time (\bcirc{2}), which enables the BL to sink current through the NAND string depending on the cell's \vth level (i.e., the BL can sink current when \vth $<$ \vref{}).\footnote{To ensure that only the target cell's \vth level affects the current through the BL, the gate voltage of all other cells in the same NAND string is set to $\text{V}_\text{PASS}$ ($>$ 6V), which is much higher than the highest \vth level of any flash cell~\cite{cai-hpca-2015, cai-procieee-2017, cai-insidessd-2018}}
The chip then enters the evaluation phase (\textbf{II} in \fig{\ref{fig:sensing}}) in which it disconnects the BL from $\text{V}_\text{PRE}$ (\bcirc{3}) and enables the sense amplifier (SA) (\bcirc{4}).
If the target cell had been programmed (i.e., \vth $>$ \vref{}), the capacitance of $\text{C}_\text{SO}$ hardly changes as the BL cannot sink current.
In contrast, if the cell had been erased, charge in $\text{C}_\text{SO}$ quickly flows through the BL, which rapidly decreases the SO-node voltage below the SA's reference voltage $\text{V}_\text{SR}$.
Finally, the chip discharges the BL (\textbf{III} in \fig{\ref{fig:sensing}}) to return to the initial state for future operations (\bcirc{5} and \bcirc{6}).
As a result, the chip-level read latency \tr can be expressed as follows: 
\begin{equation}
    \tr = N_\text{SENSE}\times(\tpre + \teval + \tdisch)
    \label{eq:tr}
\end{equation}
where $N_\text{SENSE}$ is the number of sensing times required to read a page, and \tpre, \teval, and \tdisch are the \emph{timing parameters} that define the latency for the precharge, evaluation, and discharge phases, respectively.
In SLC NAND flash memory, $N_\text{SENSE}$~$=$~1 because there are only two \vth states, while $N_\text{SENSE}$ increases up to 3 in TLC NAND flash memory to identify a specific \vth state out of eight ($=2^3$) different \vth states~\cite{cai-procieee-2017, cai-insidessd-2018}.

Manufacturers carefully decide the three timing parameters to ensure correct operation.
For example, if \tpre is too short to fully charge the BL and $\text{C}_\text{SO}$, $\text{V}_\text{SO}$ can be lower than $\text{V}_\text{SR}$ in the evaluation phase even when the target cell is programmed.
A too-short \tdisch can also lead to raw bit errors by leaving some BLs partially charged.
Since it takes more time to stabilize all BLs when there are some partially-charged BLs compared to when all BLs are fully discharged,
the next precharge phase would likely fail to properly set all BLs to $\text{V}_\text{PRE}$ within the \tpre{} latency.

\subsection{Reliability Problems in NAND Flash\label{subsec:reliability_prob}}
\begin{figure}[!h]
	\centering
	\vspace{-.5em}
	\includegraphics[width=1\linewidth]{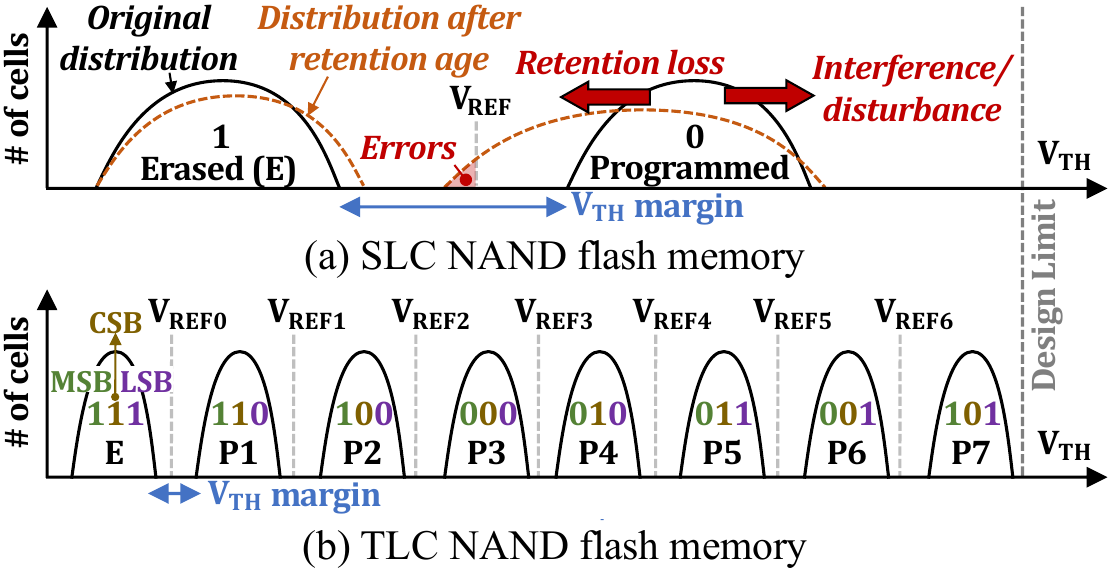}
	\vspace{-2.5em}
	\caption{\vth distribution of NAND flash memory cells.}
	\label{fig:vth_dist}
	\vspace{-1em}
\end{figure}

In NAND flash memory, a variety of sources including program interference~\cite{cai-hpca-2017, cai-iccd-2013, park-dac-2016, kim-dac-2017}, read disturbance~\cite{cai-dsn-2015, ha-ieeetcad-2015}, and data retention loss~\cite{cai-hpca-2015, cai-inteltechj-2013, cai-iccd-2012, luo-sigmetrics-2018} introduce bit errors in stored data~\cite{cai-insidessd-2018, cai-procieee-2017, cai-date-2012, cai-date-2013}.
\fig{\ref{fig:vth_dist}(a)} shows the \vth distribution of a WL in SLC NAND flash memory and how it is affected by various error sources. Reading or programming a flash cell (i.e., WL) slightly increases the \vth level of other \emph{cells} (in other WLs) in the same block by unintentionally injecting electrons to their charge traps (i.e., read disturbance and program interference).
A flash cell also leaks electrons in its charge trap over time (i.e., retention loss), which decreases the cell's \vth level.
If a cell's \vth level moves beyond the \vref{} value, a bit error occurs as the cell's data is sensed to be different from the data originally programmed into it.
Prior works show that retention loss is the dominant source of errors in 3D NAND flash memory~\cite{cai-insidessd-2018, luo-hpca-2018, luo-sigmetrics-2018, shim-micro-2019, kim-asplos-2020}.
Compared to 2x-nm planar NAND flash memory, 3D NAND flash memory experiences 40\% less program interference and 96.7\% weaker read disturbance while it suffers from a larger number of retention errors that occur faster~\cite{luo-sigmetrics-2018}.

A flash cell becomes more susceptible to errors as it experiences more program and erase (P/E) cycles~\cite{cai-date-2013, jeong-fast-2014, luo-ieeejsac-2016}.
The high voltage applied to the WL and substrate during program and erase operations damages the flash cell's tunnel oxide, which causes its charge trap to more easily get/leak electrons.
After a certain number of P/E cycles, a flash cell is \emph{worn out} (i.e., it cannot be used any longer) as it cannot retain its stored data for a required \emph{retention age}, i.e., how long data is stored after it is programmed (e.g., 1 year~\cite{jedec-2016, cox-fms-2018}).

The multi-level cell (MLC) technique aggravates the reliability problems in NAND flash memory.
As shown in \fig{\ref{fig:vth_dist}(b)}, TLC NAND flash memory stores three bits in a single cell using eight (i.e., $2^3$) different \vth states (i.e., levels).
To pack more \vth states within the same voltage window, MLC NAND flash memory inevitably \emph{narrows the margin} between adjacent \vth states, which increases the probability that a cell programmed into a particular \vth state is misread as belonging to an adjacent \vth state.  

\subsection{Reliability Management in NAND Flash\label{subsec:reliability_mgmt}}

\head{Error-correcting Codes (ECC)}
To guarantee the reliability of stored data, it is common practice in modern SSDs to employ error-correcting codes (ECC).
ECC can detect and correct bit errors within a unit of data, called a codeword, by storing redundant bits (i.e., ECC parity) into the codeword.
To address significant reliability degradation in modern NAND flash memory, a modern SSD typically adopts sophisticated ECC, such as Bose-Chauduri-Hocquenghem (BCH)~\cite{bose-infocntl-1960} and low-density parity-check (LDPC)~\cite{gallager-iretit-1962} codes, which can correct up to several tens of raw bit errors within a codeword (e.g., 72 bit errors per 1-KiB codeword~\cite{micron-flyer-2016}).

\head{Read-Retry Operation}
As modern NAND flash memory becomes more susceptible to errors, it is challenging even for strong ECC to guarantee the reliability of stored data: a page's \emph{raw bit-error rate} (RBER, the fraction of error bits in a codeword before ECC) quickly increases beyond the \emph{ECC capability} (i.e., the error-correction capability of ECC, defined as the number of error bits correctable per codeword).
This significantly degrades the lifetime of NAND flash memory since a block's lifetime is determined by the number of P/E cycles that can be performed until the block can retain the RBER lower than the ECC  capability for a minimum retention requirement~\cite{jedec-2016, cox-fms-2018, cai-procieee-2017}.
Using more sophisticated ECC (with higher ECC capability) can mitigate the lifetime degradation, but it also introduces significant area and latency overheads~\cite{cai-iccd-2012, cai-procieee-2017, cai-insidessd-2018}.

\begin{figure}[b]
    \centering
    \vspace{-1.5em}
    \includegraphics[width=\linewidth]{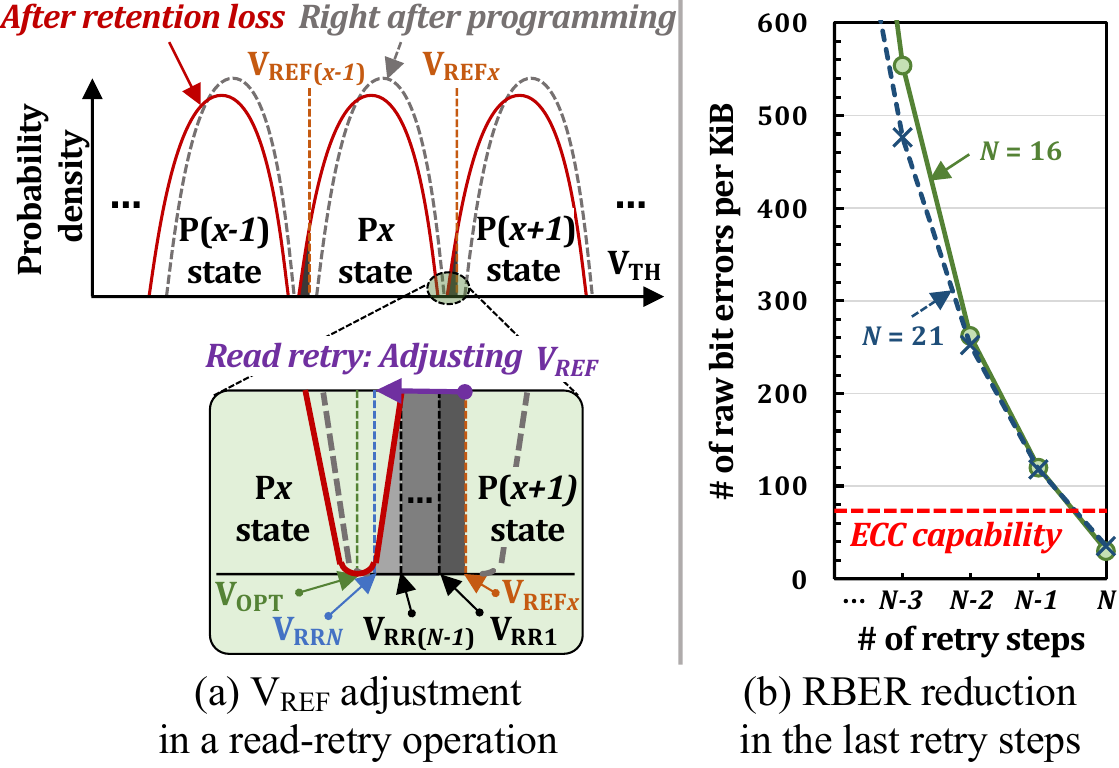}
    \vspace{-2.5em}
    \caption{RBER reduction via the read-retry mechanism.}
    \label{fig:rr}
\end{figure}

To address this, a modern SSD commonly adopts a mechanism called \emph{read-retry}~\cite{cai-date-2012, cai-date-2013, fukami-di-2017, lee-uspatent-2014, shim-micro-2019, yang-fms-2011}.
\fig{\ref{fig:rr}} shows how read-retry reduces a page's RBER to be lower than the ECC capability.
As shown in \fig{\ref{fig:rr}(a)}, retention loss \emph{shifts} and \emph{widens} the \vth distribution of each state, increasing the number of flash cells whose \vth level moves beyond the corresponding \vref{} value (e.g., \vref{x} for the P($x+1$) state).
When the number of such cells becomes higher than the ECC capability, a \emph{read failure} occurs, and the SSD controller invokes a read-retry operation for the page.
The read-retry operation reads the page \emph{again} with \emph{different} \vref{} values (e.g., \vrr{i} at the $i$-th read-retry step in \fig{\ref{fig:rr}(a)}), which decreases the number of cells misread as belonging to another \vth state. 
The controller performs further retry steps until it either successfully reads the page without uncorrectable errors or fails to reduce the page's RBER to a value lower than the ECC capability even after trying all of the \vref{} values that are available to the mechanism.

How to adjust the \vref{} values is the most critical design choice in the read-retry mechanism. 
The \vth distribution of each state is \emph{very narrow} in modern MLC NAND flash memory to store $m$ bits per cell using $2^{m}$ \vth states (e.g., 16 \vth states in QLC NAND flash memory).
This causes the page's RBER to be extremely sensitive to the distance of the \vref{} value from the \emph{optimal read-reference voltage \vopt}.
In \fig{\ref{fig:rr}(a)}, for example, we can see that \vrr{(N-1)} leads to a significantly larger number of bit errors (i.e., a wider gray area before \vrr{(N-1)}) compared to \vrr{N}, which is closer to \vopt.

Through extensive profiling of NAND flash chips, manufacturers provide sets of \vref{} values used for a read-retry operation which guarantee the \vref{} values in the final retry step to be substantially close to \vopt. 
\fig{\ref{fig:rr}(b)} shows how two pages' RBER values change in the \emph{last four} retry steps when reading the two pages requires 16 and 21 retry steps, respectively.
We measure the RBER values from real 3D TLC NAND flash chips (see \sect{\ref{sec:prof_method} for detailed description of our infrastructure and methodology)}. As shown in \fig{\ref{fig:rr}(b)}, each page's RBER \emph{drastically decreases in the final (i.e., $N$-th) retry step} due to the use of \emph{near-optimal} \vref{} values, enabling the correct reading of the page.

%Although the read-retry mechanism is essential to
The read-retry mechanism is essential improving reliability and enhancing SSD lifetime, but a read-retry operation can significantly degrade SSD performance due to \emph{multiple} retry steps it causes.
In general, the page-read latency \tread can be formulated as follows:
\begin{equation}
    \tread = \tr + \tdma + \tecc + \trr
    \label{eq:tread}
\end{equation}
where \tr, \tdma, \tecc, and \trr are the latencies of sensing the page data (Equation~\eqref{eq:tr}), transferring the sensed data from the chip to the SSD controller, decoding the data with the ECC engine, and performing a read-retry operation, respectively.
When a page read requires \nrr ($\geq 0$) retry steps, \trr can be expressed as follows~\cite{cai-hpca-2015, shim-micro-2019}:
\begin{equation}
    \trr = N_\text{RR} \times (\tr + \tdma + \tecc).
    \label{eq:trr}
\end{equation}
Since a read-retry operation increases \tread linearly with \nrr, it can significantly degrade SSD performance.

\vspace{-.3em}
\section{Motivation\label{sec:motivation}}
\vspace{-.3em}
In this section, we 1) present read-retry characteristics of modern NAND flash memory, and 2) introduce two new opportunities for reducing the read-retry latency. 

\vspace{-.3em}
\subsection{Read-Retry in Modern NAND Flash\label{subsec:rr_behavior}}
\begin{figure}[!t]
    \centering
    \includegraphics[width=\linewidth]{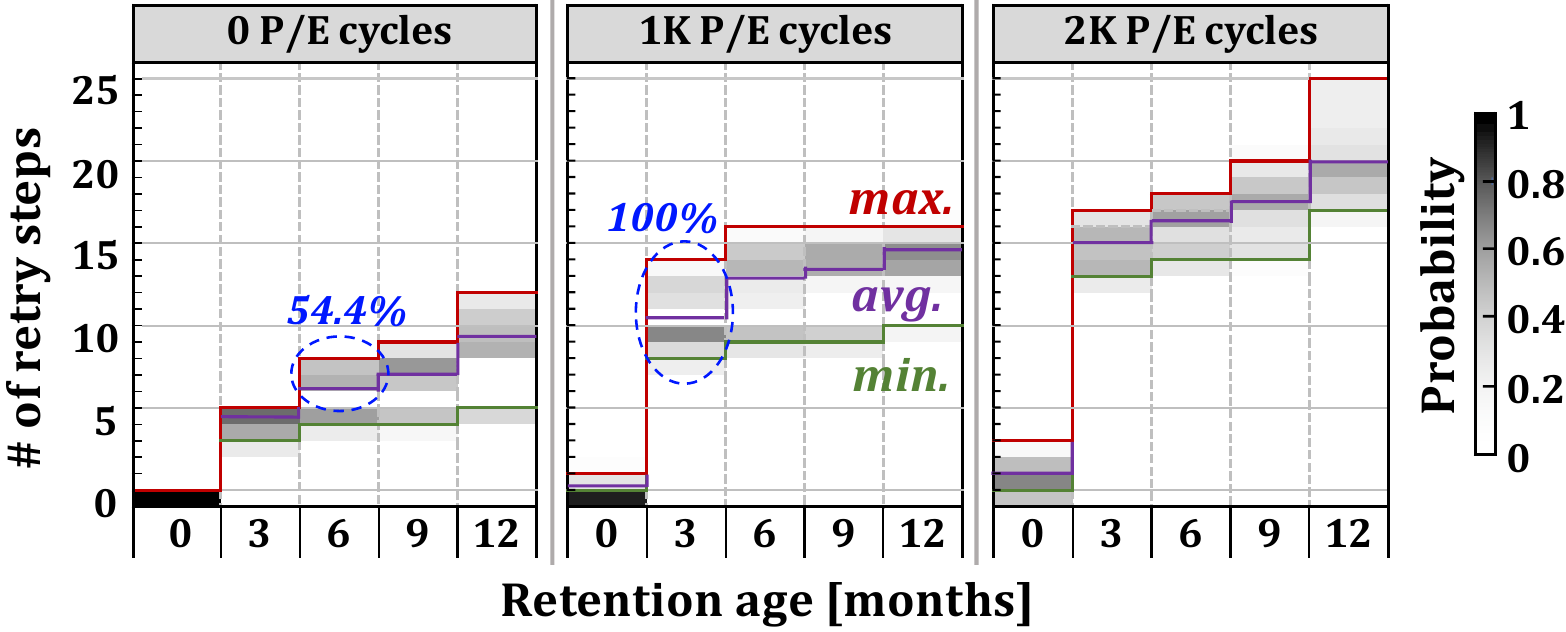}
    \vspace{-2.5em}
    \caption{Read-retry characteristics of 160 3D TLC NAND flash memory chips under different operating conditions.}
    \label{fig:rr_dist}
    \vspace{-1.5em}
\end{figure}

\vspace{-.3em}
To understand how many read-retry operations occur in modern NAND flash memory and how frequently they occur, we characterize 160 real 3D TLC NAND flash chips under different operating conditions (see \sect{\ref{sec:prof_method}} for detailed description of our infrastructure and methodology).
We measure the number of read-retry steps for more than $10^7$ pages that are randomly selected from the 160 NAND flash chips, under different operating conditions.
\fig{\ref{fig:rr_dist}} shows the probability of occurrence of different numbers of retry steps (in gray scale) for different P/E-cycle counts and \emph{retention ages}.
A box at ($x$, $y$) represents the probability that a read requires a read-retry operation with $y$ retry steps under $x$-month retention age.
\fig{\ref{fig:rr_dist}} plots the probability under three different P/E-cycle counts, 0 (left), 1K (center), and 2K (right).

We make two observations from the results. 
First, a page read introduces a significant number of retry steps especially when the page experiences more P/E cycling and/or has a longer retention age.
While a \emph{fresh} page (i.e., with no P/E cycling and 0 retention age) can be read without a read-retry, the average number of retry steps significantly increases to 19.9 under a 1-year retention age at 2K P/E cycles, which in turn increases \tread by 21$\times$ on average.
Second, a read-retry occurs very frequently even under modest operating conditions, introducing a number of retry steps.
\fig{\ref{fig:rr_dist}} shows that 54.4\% of reads incur at least seven retry steps under a 6-month retention age even when the pages have \emph{never} experienced P/E cycling (the dot-circle in the left plot).
At 1K P/E cycles, at least eight read-retry steps are needed to read a page only after a 3-month retention age (the dot-circle in the center plot of \fig{\ref{fig:rr_dist}}).
This means that the performance degradation due to read-retry operations can be significant not only under worst-case conditions but also under the common case.

Our characterization results clearly show the importance of mitigating the read-retry overhead.
Prior works propose several techniques that reduce the \emph{number} of retry steps~\cite{cai-hpca-2015, cai-iccd-2013, luo-ieeejsac-2016, luo-hpca-2018, luo-sigmetrics-2018, nie-dac-2020, shim-micro-2019}, but read-retry is difficult to \emph{completely avoid} in modern SSDs as \vopt quickly and significantly changes over time. 
For example, an existing technique can reduce the average number of read-retry steps by about 70\% under a 1-year retention age at 2K P/E cycles, but for \emph{every} page read, it requires at least three retry steps~\cite{shim-micro-2019}.

\subsection{Optimization Opportunities for Read-Retry\label{subsec:opportunities}}
We identify two new opportunities to reduce \trr by exploiting two advanced architectural features in modern SSDs: 1) the \cread command and 2) the strong ECC engine.

%\subsubsection{Exploiting the \cread Feature.\label{sssec:cread}}
\subsubsection{\textbf{Exploiting the \cread Feature.}\label{sssec:cread}}
Modern NAND\linebreak flash memory supports an advanced command called \cread \cite{leong-uspatent-2008, macronix-technote-2013, micron-technote-2004, micron-datasheet-2009, samsung-datasheet-2009, toshiba-datasheet-2012} that can effectively reduce \tread by pipelining consecutive read requests.\footnote{The \cread command requires an additional \emph{cache (i.e., page buffer)} in the NAND flash chip to store sensed data while transferring the previously-sensed data to the SSD controller.}
Early generations of NAND flash memory support the \cread feature only for \emph{sequential reads} (i.e., only when the target page of an incoming read is \emph{physically next} to the currently accessed)~\cite{micron-technote-2004}.
However, to improve the random-read performance, which is critical in popular applications~\cite{liu-asplos-2019, park-nvmsa-2018}, such as key-value stores~\cite{atikoglu-sigmetrics-2012} and graph analytics~\cite{zheng-fast-2015},
manufacturers including Samsung, Micron, and Toshiba have extended the \cread  command to support \emph{any} consecutive page reads regardless of the locations of the pages to be read~\cite{leong-uspatent-2008, macronix-technote-2013, micron-datasheet-2009, samsung-datasheet-2009, toshiba-datasheet-2012}.

\begin{figure}[!t]
    \centering
    \includegraphics[width=.95\linewidth]{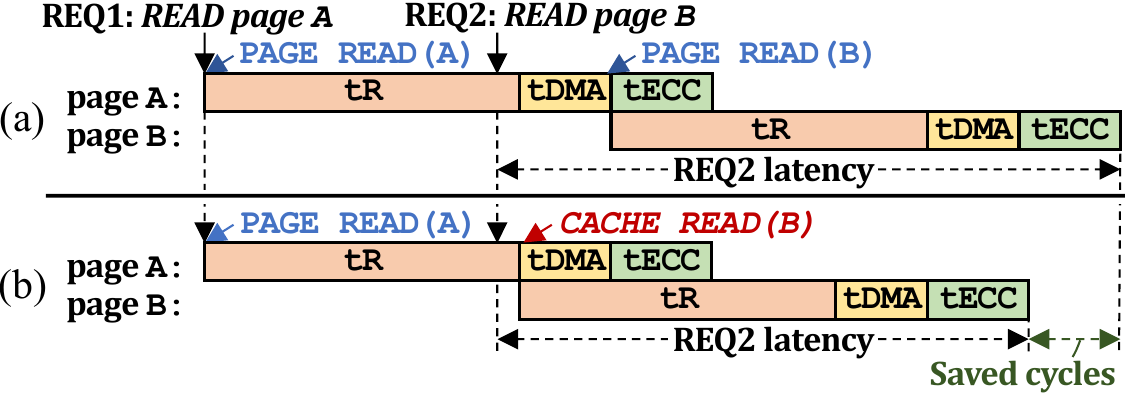}
    \vspace{-1.5em}
    \caption{Comparison of (a) basic \texttt{PAGE}~\texttt{READ} command and (b) \cread command (see page~\texttt{B} in each figure).}
    \label{fig:cread}
    \vspace{-1em}
\end{figure}

\fig{\ref{fig:cread}} shows how an SSD controller reduces the latency of a page read using the \cread command.
As shown in \fig{\ref{fig:cread}(a)}, with the basic \texttt{PAGE}~\texttt{READ} command, an SSD controller can start reading page~\texttt{B} \emph{only after} finishing the data transfer of page~\texttt{A}. (The data of page \texttt{A} is decoded by the ECC engine dedicated to the channel~\cite{cai-procieee-2017}, so the SSD controller can concurrently perform sensing of page~\texttt{B} with ECC decoding of page~\texttt{A}.)
In contrast, as shown in \fig{\ref{fig:cread}(b)}, the SSD controller can issue a \cread command for page~\texttt{B} \emph{before} starting the data transfer of page~\texttt{A} so that the chip can concurrently perform both the data transfer of page~\texttt{A} and sensing of page~\texttt{B}.
Since each retry step of a read-retry operation is effectively the same as a regular page read, we can also perform consecutive retry steps in a pipelined manner via the \cread command, which in turn reduces the total execution time of a read-retry operation.    

%\subsubsection{Exploiting Large ECC-Capability Margin.\label{sssec:ecc_cap}}
\subsubsection{\textbf{Exploiting Large ECC-Capability Margin.}\label{sssec:ecc_cap}}
We find that there would be \emph{a large ECC-capability margin\footnote{\emph{ECC-Capability Margin $=$ Maximum Number of Raw Bit Errors a Given ECC can Correct per Codeword $-$ Number of Raw Bit Errors Present in a Codeword.}} when a read-retry occurs}.
This may sound contradictory as a read-retry occurs only when the page's RBER exceeds the ECC capability, i.e., when there is no ECC-capability margin.
However, when a read-retry operation succeeds, the page is eventually read \emph{without} any uncorrectable errors, which means that there exists \emph{a positive} ECC-capability margin in the final retry step if it succeeds.
We hypothesize that the ECC-capability margin is large due to two reasons.
First, as explained, a modern SSD uses \emph{strong} ECC that can correct several tens of raw bit errors in a codeword.
Second, in the final retry step, the page can be read by using \emph{near-optimal} \vref{} values that drastically decrease the page's RBER as explained in \sect{\ref{subsec:reliability_mgmt}}.

If we can empirically demonstrate and methodically leverage the large ECC-capability margin in the final retry step to reduce the \emph{page-sensing latency \tr}, doing so allows us to reduce \trr considerably.
This is because \tr is the dominant factor in \trr especially when we use the \cread command in a read-retry operation.
Although reducing \tr may increase the page's RBER as explained in \sect{\ref{subsec:nand_opr}}, we can \emph{safely} reduce \tr for a read-retry operation as long as the number of additional bit errors introduced by the reduced \tr is lower than the large ECC-capability margin in the final retry step.
We hypothesize that this is the common case since manufacturers \emph{pessimistically} set the timing parameters to cover for the \emph{worst-case} operating conditions and process variation~\cite{chang-sigmetrics-2016, kim-iccd-2018, lee-sigmetrics-2017, lee-hpca-2015, mutlu-superfri-2015, mutlu-imw-2013}.
For example, an outlier BL can have much higher capacitance than other BLs due to its geometry (e.g., thick wire, narrow contacts, and high parasitic capacitance), which significantly increases the time for the BL to be fully charged.
Even if the fraction of such BLs may be very low, eliminating all of them from a chip either requires extreme effort or leads to a significant loss in chip yield.
Consequently, such outlier BLs dictate \tr, even though most BLs can correctly operate with reduced \tr.

\vspace{-.1em}
\section{Characterization Methodology\label{sec:prof_method}}
\vspace{-.1em}
To test our hypothesis in \sect{\ref{sssec:ecc_cap}}, we characterize 1) the ECC-capability margin in the final retry step and 2) the reliability impact of \tr reduction, using 160 real 3D TLC NAND flash chips.

\head{Infrastructure}
We use an FPGA-based testing platform that contains a custom flash controller and a temperature controller.
The flash controller allows us to access a NAND flash chip using all the commands implemented in the chip.
It supports not only basic read/program/erase operations, but also dynamic change of timing parameters for a read by using the \texttt{SET}~\texttt{FEATURE} command~\cite{onfi-2020}.
The temperature controller maintains the temperature of a NAND flash chip within \degreec{\pm1} of the target temperature.
This allows us to test a NAND flash chip under different operating temperatures (i.e., the temperature when a page is read or programmed) and accelerate retention loss based on Arrhenius's Law~\cite{arrhenius-zpc-1889} (e.g., 13 hours at \degreec{85} $\approx$ 1 year at \degreec{30}).
We characterize 160 48-layer 3D TLC NAND flash chips in which $\langle$\tpre, \teval, \texttt{tDISCH}$\rangle =$~$\langle24$~\usec, 5~\usec, 10~$\mu$s$\rangle$~(i.e., \tpre:\teval:\tdisch$\approx$~5:1:2) by default.

\head{Methodology}
To minimize the potential distortions in our characterization results, we randomly select 120 blocks from each of the 160 3D NAND flash chips at different physical block locations and perform read tests for every page in each selected block.
We test a total of 3,686,400 WLs (11,059,200 pages) to obtain statistically significant experimental results.
Unless specified otherwise, we report a representative (i.e., maximum and/or average) value across all the tested pages from the 160 chips.
For a \emph{read test} of a page, we first read the target page with default read-timing parameters and measure the page's RBER.
When a read failure occurs, we perform a read-retry operation while measuring the page's RBER in each retry step. 
Then, using the same \vref{} values used in the final retry step, we repeat reading of the page and measure its RBER while reducing read-timing parameters, in order to evaluate the impact of reducing the timing parameters on the RBER in the final retry step.

We perform read tests while varying the P/E-cycle count, retention age, and operating temperature, all of which are shown to significantly affect a flash cell's error behavior~\cite{cai-date-2012, cai-date-2013, cai-hpca-2015, cai-iccd-2013, cai-inteltechj-2013, cai-procieee-2017, cai-insidessd-2018, luo-hpca-2018, luo-ieeejsac-2016, luo-sigmetrics-2018, shim-micro-2019}.
We follow the test procedures of the JEDEC industry standard~\cite{jedec-2016} in each read test.
To increase the P/E-cycle count of a block, we repeat the cycle of 1) programming every page in the block with random data\footnote{Although a page's RBER has data-pattern dependence~\cite{cai-date-2012, cai-procieee-2017, cai-iccd-2012, luo-sigmetrics-2018}, we use random data because modern SSDs commonly use a \emph{data randomizer}~\cite{cha-etri-2013, cai-procieee-2017, kim-ieeejssc-2012, lin-uspatent-2012} to avoid the worst-case data patterns that may cause unexpected read failures.} and 2) erasing the block.
For each target P/E-cycle count, we test each page while varying the retention age and operating temperature by using the temperature controller. 

\vspace{-.3em}
\section{Characterization Results\label{sec:prof_results}}
\vspace{-.3em}
We present and analyze our real-device characterization results on 1) the ECC-capability margin in the final retry step and 2) the reliability impact of reducing read-timing parameters, collected across 160 3D TLC NAND flash chips. 

\vspace{-.3em}
\subsection{ECC-Capability Margin in Final Retry Step\label{subsec:margin}}
\begin{figure}[!h]
    \centering
    \vspace{-.5em}
    \includegraphics[width=\linewidth]{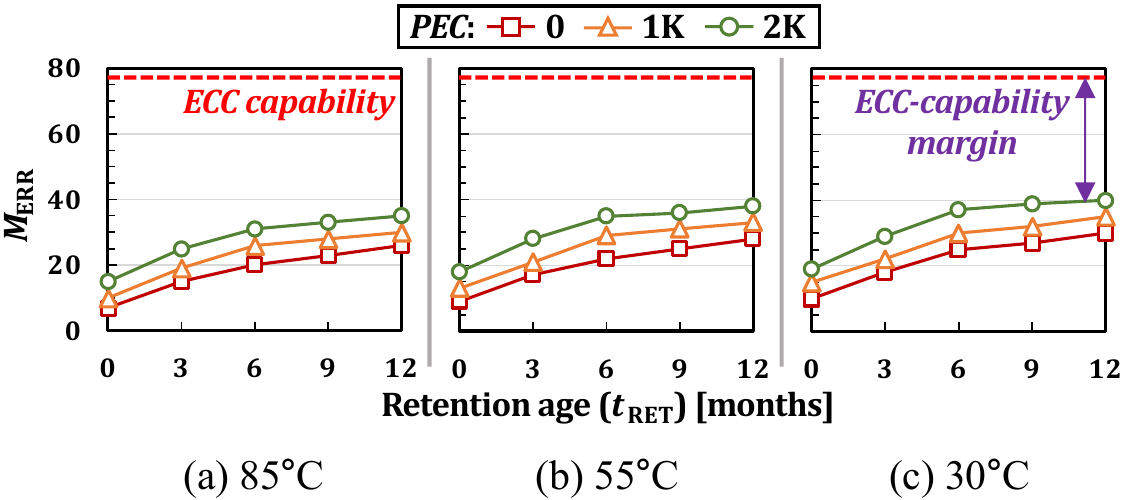}
    \vspace{-2.5em}
    \caption{ECC-capability margin in the final read-retry step.}
    \label{fig:margin}
    \vspace{-1em}
\end{figure}

\vspace{-.3em}
\fig{\ref{fig:margin}} depicts \nerri{\pec}{\tret}, i.e., the maximum number of bit errors per 1-KiB data in the final read-retry step
under different P/E-cycle counts (\pec) and retention ages (\tret, unit: months)\footnote{A flash cell's retention loss is significantly affected by ambient temperature. We show the \emph{effective} retention age at \degreec{30}, following an industrial standard that specifies the retention requirements of NAND flash-based SSDs~\cite{jedec-2016}.}, at three different operating temperatures: (a) \degreec{85}, (b) \degreec{55}, and (c) \degreec{30}.
We also plot the ECC capability at 72 errors per 1 KiB.
We make three key observations.
First, there is a large ECC-capability margin in the final retry step \emph{even under the worst-case operating conditions} prescribed by manufacturers (e.g., a 1-year retention age~\cite{cox-fms-2018} at 1.5K P/E cycles~\cite{micron-flyer-2016}).
We observe that even \nerri{2K}{12} at \degreec{30} is quite low, leaving a margin as large as 44.4\% of the ECC capability.
This shows that, although strong ECC is an inevitable choice for a modern SSD, its high ECC capability is largely underutilized when a read-retry eventually succeeds, due to the use of near-optimal \vref{} values in the final retry step.

Second, the ECC-capability margin decreases as the page experiences more P/E cycling and longer retention age (e.g., \nerri{0}{3} $=$~15 while \nerri{1K}{12}$=$~30 at \degreec{85}).
This is due to the inherent error characteristics of NAND flash memory.
In fact, unlike what is idealistically shown in \figs{\ref{fig:vth_dist} and \ref{fig:rr}},\footnote{The \vth distribution of NAND flash memory is usually described with simplified figures (similar to \figs{\ref{fig:vth_dist} and \ref{fig:rr}}) to ease understanding~\cite{jeong-fast-2014, luo-hpca-2018, luo-sigmetrics-2018, park-dac-2016, shim-micro-2019}.} two adjacent \vth states slightly overlap even right after programming a \emph{fresh} page, which makes \emph{no \vref{} value capable of achieving zero RBER} in modern NAND flash memory~\cite{cai-procieee-2017, cai-hpca-2015, luo-ieeejsac-2016, luo-sigmetrics-2018, luo-hpca-2018}.
As P/E cycling and retention age shift and widen \vth state distributions, even the optimal read-reference voltage cannot completely avoid the RBER increase. 

Third, operating temperature also affects the ECC-capability margin in the final retry step, but its impact is \emph{not} as significant as P/E cycling and retention age.
Compared to \degreec{85}, \nerr at \degreec{30} and \degreec{55} is higher by 5 and 3 errors, respectively, all other conditions being equal. 
In 3D NAND flash memory, an electron's mobility in the poly-type channel, which decreases with operating temperature, is the dominant factor affecting the cell current through the BL~\cite{arya-thesis-2012}. 
Since an erased cell might be recognized as programmed due to reduced current, \nerr slightly increases as operating temperature reduces.
We observe the same relationship in all the tested chips (i.e., the lower the temperature, the higher the page's RBER).

We draw two conclusions based on our observations.
First, we can use the large ECC-capability margin in the final retry step to reduce \trr, unless reduction of read-timing parameters significantly increases the page's RBER.
Second, the ECC-capability margin highly depends on operating conditions, so we should carefully decide the reduction amount considering the current operating conditions.

\subsection{Reliability Impact of Reducing Read-Timing Parameters \label{subsec:reduce_rt}}
We first present the effect of reducing individual read-timing parameters (\sect{\ref{sssec:indv_rt}}).
We then show the effect of reducing multiple timing parameters simultaneously (\sect{\ref{sssec:mult_rt}}) and summarize our characterization results with the final timing parameters we decide for reliable \trr reduction (\sect{\ref{sssec:rrr}}). 

%\subsubsection{Reduction of Individual Parameters.\label{sssec:indv_rt}}
\subsubsection{\textbf{Reduction of Individual Parameters.}\label{sssec:indv_rt}}
\begin{figure}[!b]
    \centering
    \includegraphics[width=\linewidth]{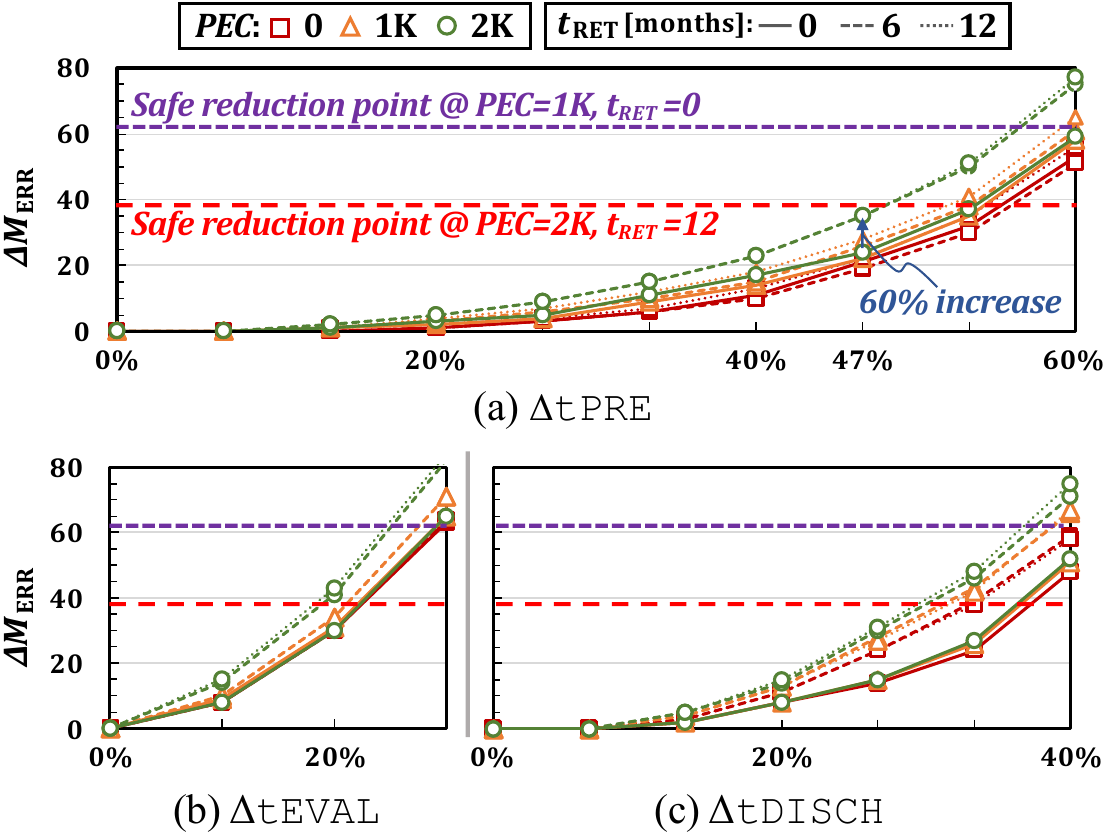}
    \vspace{-2em}
    \caption{Effect of reducing each read-timing parameter.}
    \label{fig:indv}
\end{figure}

We first evaluate the effect of reducing individual read-timing parameters under different operating conditions.
\fig{\ref{fig:indv}} shows \derr, the maximum \emph{increase} of raw bit errors per 1-KiB data when we read a page at \degreec{85}\footnote{We show the effect of operating temperature in \sect{\ref{sssec:rrr}}.} with reduced (a) \tpre, (b) \teval, or (c) \tdisch, compared to when using the default value.

We make three observations from the results.
First, it is possible to safely reduce read-timing parameters for optimizing the read-retry latency. 
Even under a 1-year retention age at 2K P/E cycles (where \nerr$=35$), we can safely reduce \tpre, \teval, and \tdisch by 47\%, 10\%, and 27\%, respectively.
Second, reduction in \teval or \tdisch leads to faster increase in \nerr compared \tpre, which implies that manufacturers set the default \tpre{} \emph{more pessimistically} than the other parameters.
As explained in \sect{\ref{subsec:nand_opr}, the precharge phase needs to \emph{stabilize} \emph{every BL} at a certain voltage level ($\text{V}_\text{PRE}$), which requires a large timing margin in \tpre for outlier BLs.
On the other hand, the discharge phase requires a relatively-small timing margin in \tdisch compared to \tpre because it only pulls out $\text{V}_\text{PRE}$ from BLs.}
Third, P/E cycling and retention age also affect the increase in bit errors due to reduced read-timing parameters as well as the ECC-capability margin in the final retry step.
In particular, we observe non-trivial impact of retention age on \derr.
When reducing \tpre by 47\%, for example, \derri{2K}{12} is 60\% higher than \derri{2K}{0} (i.e., a 1-year retention age increases \derr by 60\% at 2K P/E cycles) as shown in \fig{\ref{fig:indv}(a)}.

We draw two conclusions based on our observations.
First, we can significantly reduce \tr in a read-retry operation \emph{even under the worst-case operating conditions} prescribed by manufacturers.
Our results demonstrate that \tpre can be safely reduced by \emph{at least} 40\% under every tested condition, which leads to a 25\% reduction in \tr. 
Second, it is \emph{very cost-ineffective} to reduce \teval.
Reducing \teval by 20\% introduces 30 additional bit errors (i.e., 41.7\% of the ECC capability) \emph{even for a fresh page}.
This significantly decreases the chance to reduce the other parameters while achieving only 2.5\%~\tr reduction due to the low contribution of \teval{} to \tr (1/8 \tr only).
Therefore, we decide to exclude \teval{} from our later analyses.

\subsubsection{\textbf{Reduction of Multiple Parameters.}\label{sssec:mult_rt}}
\begin{figure}[b]
    \centering
    \vspace{-1em}
    \includegraphics[width=\linewidth]{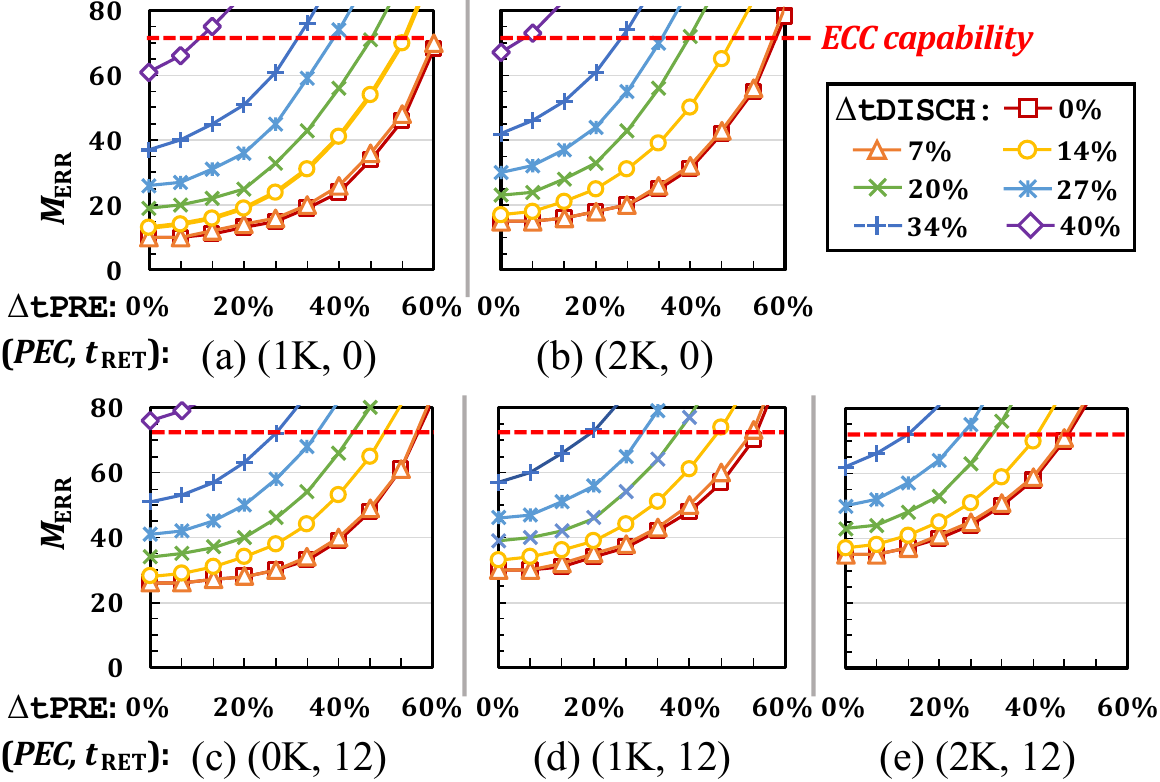}
    \vspace{-2em}
    \caption{Effect of reducing multiple read-timing parameters, \tpre and \tdisch, under different P/E-cycle counts (\emph{PEC}) and retention ages ($t_\text{RET}$, unit: months).}
    \label{fig:mult}
\end{figure}

%\subsubsection{Reduction of Multiple Parameters.\label{sssec:mult_rt}}
Although our results of the previous experiments promise a great opportunity for reducing each of \tpre and \tdisch alone, reducing one may decrease the chance of reducing the other (because the discharge phase of a read affects the precharge phase of the next read as explained in \sect{\ref{subsec:nand_opr}}).
To identify the potential for reducing both timing parameters simultaneously, we test all possible combinations of (\tpre, \tdisch) values while reducing \tpre by up to 60\% and \tdisch by up to 40\%.
\fig{\ref{fig:mult}} plots \nerri{\pec}{\tret}, the maximum number of bit errors per 1-KiB data in the final retry step when we read test pages while reducing \tpre and \tdisch simultaneously under five different operating conditions.

We make three key observations based on the results.
First, reducing the two timing parameters simultaneously introduces more additional bit errors than reducing each parameter individually.
For example, as shown in \figs{\ref{fig:indv}(a) and \ref{fig:indv}(c)}, when we reduce \tpre by 54\% and \tdisch by 20\% individually, \derri{1K}{0} is 35 and 8, respectively.
Unfortunately, simultaneous reduction of the two timing parameters increases \nerr far beyond the ECC capability; see the green line with marker $\times$ in \fig{\ref{fig:mult}(a)} (i.e., $\Delta$\tdisch$=$20\%), which is outside the plot at $\Delta$\tpre$=54$\%.
Second, it is more beneficial to reduce \tpre than to reduce \tdisch in most cases.
\nerr is smaller when $\langle\Delta$\tpre, $\Delta$\tdisch$\rangle$ = $\langle x\%, y\%\rangle$ compared to when $\langle\Delta$\tpre, $\Delta$\tdisch$\rangle$ = $\langle y\%, x\%\rangle$ for most values of $x$ and $y$.
Third, despite the higher reliability impact of \tdisch over \tpre{} (discussed in \sect{\ref{sssec:indv_rt}}), reducing \tdisch by 7\% hardly increases the number of bit errors (by 4 at most) under every operating condition.  

Based on our observations, we conclude that it is effective to use the ECC-capability margin in the final retry step for \emph{only reducing \tpre}.
Although the increase in additional bit errors from reducing \tdisch by 7\% is quite low (up to 4 additional bit errors), the cost of doing so is larger than the benefit: considering that the fraction of \tdisch in \tr is only 25\%, a 7\% reduction in \tdisch merely reduces \tr by 1.75\% (i.e., $0.07\times0.25$), while its cost could be up to 5.6\% of the ECC capability (i.e., up to 4 additional bit errors under the ECC capability of 72 errors per 1 KiB).

%\subsubsection{Reliable Reduction of \tpre\label{sssec:rrr}}
\subsubsection{\textbf{Reliable Reduction of \tpre}\label{sssec:rrr}}
\begin{figure}[h]
    \centering
    \includegraphics[width=\linewidth]{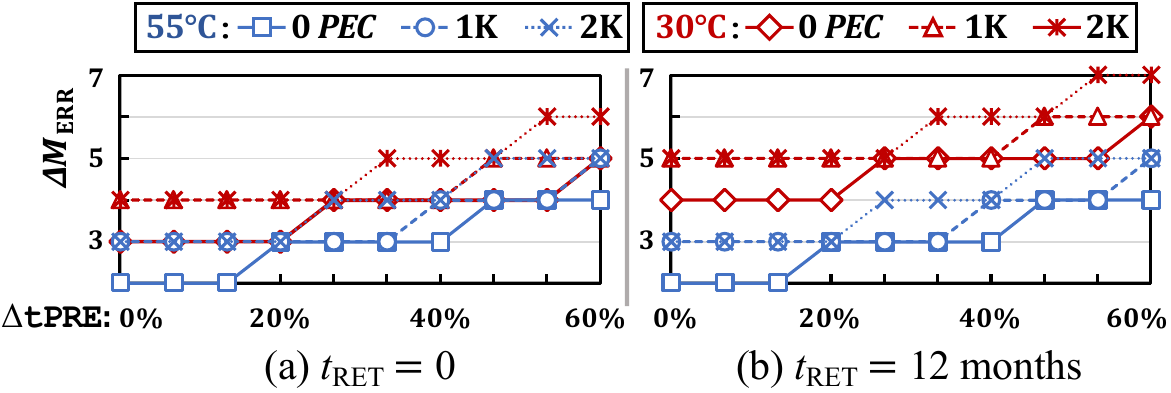}
    \vspace{-2em}
    \caption{Effect of operating temperature on the number of additional errors due to \tpre reduction.}
    \label{fig:temp}
    \vspace{-1em}
\end{figure}

As the final step of our characterization, we analyze the impact of operating temperature on the amount of \tpre reduction. 
\fig{\ref{fig:temp}} plots \derr, the increase in the maximum number of raw bit errors in the final retry step when a NAND flash chip operates at \degreec{30} and \degreec{55}, compared to at \degreec{85}.
We observe that operating temperature affects \derr in a similar way as it affects \nerr: the lower the operating temperature, the larger the \derr, and the temperature effect becomes more significant under a longer retention age and higher P/E-cycle count.
The increase in \derr is also small: it is only up to 7 additional bit errors even under a 1-year retention age at 2K P/E cycles.

Based on these results, we conclude that we should incorporate a safety margin into reduced \tpre to ensure that a page's RBER is lower than the ECC capability in the final retry step under varying operating temperature.
It is also possible to profile the \emph{optimal} \tpre for each combination of (\pec,~\tret,~$T$) where $T$ is the operating temperature.
However, we decide to determine a \emph{good} \tpre value by considering only \pec and \tret{}, and plan for sufficient ECC capability that can correct temperature-induced additional errors.
This is due to two reasons. 
First, the effect of operating temperature on \nerr is quite small compared to the effect of \pec and \tret, and thus operating temperature does not significantly affect the reduction in \tpre. 
When we reduce \tpre{} alone by less than 40\%, a substantial ECC-capability margin remains to correct temperature-induced additional errors under every operating condition.
Second, our decision greatly reduces profiling effort and eliminates the need to monitor a wide range of temperatures.
In particular, operating temperature may be difficult or costly to accurately measure for each retry step as it changes much more quickly than \pec and \tret.

\begin{figure}[h]
    \centering
    \includegraphics[width=\linewidth]{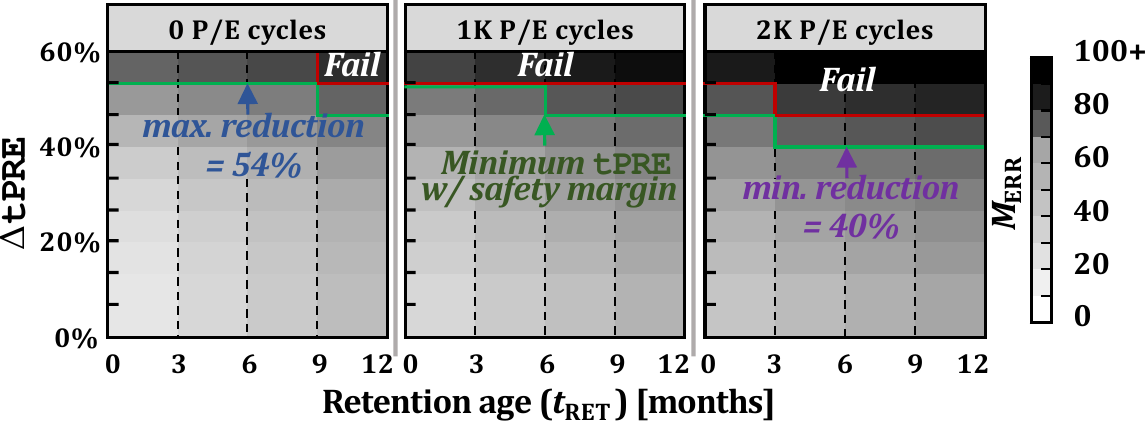}
    \vspace{-2.3em}
    \caption{Minimum \tpre for safe \trr reduction.}
    \label{fig:best}
    \vspace{-1em}
\end{figure}

\fig{\ref{fig:best}} shows the values we select for safely reducing the read-retry latency under different operating conditions.
To minimize the probability of increasing the number of retry steps for outlier pages (which could potentially be missed in the set of pages we test experimentally), we ensure that the selected \tpre value for each operating condition guarantees an ECC-capability margin of 14 bits in the final retry step (7 bits for temperature-induced errors and 7 bits for errors in outlier pages).  
We conclude that, even with the 14-bit margin, we can significantly reduce \tpre by at least 40\% (up to 54\%) under any operating condition, as shown in \fig{\ref{fig:best}}.

\vspace{-.3em}
\section{Read-retry Optimizations\label{sec:techs}}
\vspace{-.3em}
Motivated by our new experimental findings from real 3D NAND flash chips, 
we propose two new techniques that effectively reduce the read-retry latency: 1) \underline{P}ipelined \underline{R}ead-\underline{R}etry (\prr) and 2) \underline{A}daptive \underline{R}ead-\underline{R}etry (\arr).

\vspace{-.4em}
\subsection{PR$^2$: Pipelined Read-Retry\label{subsec:pr2}}
\begin{figure}[!t]
    \centering
    \includegraphics[width=\linewidth]{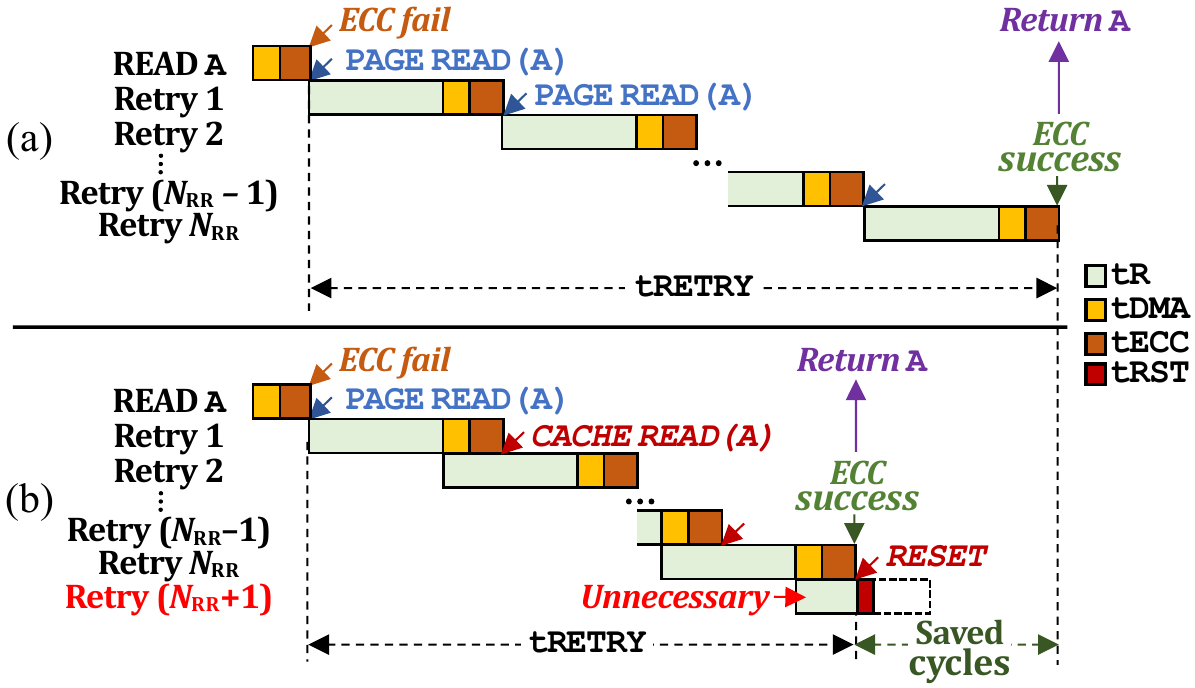}
    \vspace{-2.5em}
    \caption{Comparison of (a) regular read-retry and (b) \prr.}
    \vspace{-1em}
    \label{fig:prr}
\end{figure}

\vspace{-.3em}
\prr reduces the total execution time of a read-retry operation by pipelining consecutive retry steps using the \cread command.
\fig{\ref{fig:prr}} compares \prr with a regular read-retry operation.
As shown in \fig{\ref{fig:prr}(a)}, the existing read-retry mechanism~\cite{cai-insidessd-2018, cai-procieee-2017, cai-date-2012, cai-date-2013, fukami-di-2017, lee-uspatent-2014, shim-micro-2019, yang-fms-2011} starts a new retry step \emph{after} checking whether ECC decoding for the previous step succeeds, which places \tecc on the critical path of \trr (Equation~\eqref{eq:trr}).\footnote{The SSD controller can issue a new \pread command \emph{before starting} ECC decoding of the currently read-out page (as described in \fig{\ref{fig:cread}(a)}). However, prior works~\cite{cai-date-2012, cai-date-2013, fukami-di-2017, lee-uspatent-2014, shim-micro-2019, yang-fms-2011} assume that a new retry step starts \emph{after}} the previous retry step is completed because \emph{speculatively} starting a new retry step could delay other user requests waiting for the completion of the on-going read (and retries), if the speculative retry step ends up being not needed.
In contrast, as shown in \fig{\ref{fig:prr}(b)}, \prr starts the next retry step right after the chip completes page sensing of the current step (i.e., after \tr of the current step) by issuing a \cread command on the same target page.
Since the SSD controller can return the read page once ECC decoding succeeds, when the read request requires \nrr retry steps, \trr in \prr can be formulated as follows:
\vspace{-.1em}
\begin{equation}
    \trr = N_\text{RR} \times \tr + \tdma + \tecc.
    \label{eq:trr_new}
\end{equation}
\vspace{-.1em}
Thus, \prr reduces \trr by $($\nrr$-$1$)$$\times$$($\tdma$+$\tecc) over the regular read-retry mechanism (Equation~\eqref{eq:trr}).
Considering that many reads require multiple retry steps even under modest operating conditions as we observe in \sect{\ref{subsec:rr_behavior}} (e.g., \emph{every} read requires more than \emph{eight} retry steps under a 3-month retention age at 1K P/E cycles as shown in \fig{\ref{fig:rr_dist}}), \prr significantly improves SSD performance by effectively reducing \trr.

When a read requires \nrr retry steps, \prr speculatively starts the ($N_\text{RR}+1$)-th retry step that is unnecessary to read the page. 
This unnecessarily-started retry step could negatively affect SSD performance by delaying other operations that may exist in the request queue, waiting for the completion of the read.
\prr minimizes this potential performance penalty by using the \texttt{RESET} command that immediately terminates any on-going request that is being performed in the chip.
As described in \fig{\ref{fig:prr}(b)}, the SSD controller issues a \texttt{RESET} command as soon as ECC decoding succeeds, which takes only a few microseconds to terminate the unnecessarily-started retry step (the reset latency \texttt{tRST} = 5~\usec for a read operation~\cite{micron-datasheet-2005}).

\head{Overhead}
\prr requires no change to NAND flash chips.
It requires only slight modifications to the SSD controller or firmware to issue 1) a \cread command for each retry step immediately after page sensing of the previous step and 2) a \texttt{RESET} command as soon as ECC decoding succeeds.
The performance overhead of \prr is also small.
The unnecessarily-started retry step could delay other operations (that are waiting for the completion of the read in the request queue) only for several microseconds at most.
When the read requires more than one retry step, which is the common case (see \sect{\ref{subsec:rr_behavior}}), the latency benefit of \prr is always higher than its latency overhead.

\subsection{AR$^2$: Adaptive Read-Retry\label{subsec:ar2}}
\arr optimizes the read-retry latency by using the large ECC-\linebreak capability margin in the final retry step to reduce \tpre (and thus \tr) for \emph{every} retry step (\emph{not} only for the final step).
\arr carefully decides the \tpre reduction amount depending on the current operating conditions to avoid additional retry steps introduced by \tpre reduction.
To this end, we propose that SSD manufacturers 1) identify the best \tpre values at different operating conditions for each NAND flash chip via offline profiling of the chip and 2) incorporate the information into the SSD in the form of a simple table, i.e., \emph{Read-timing Parameter Table (RPT)}. 
The RPT stores the best profiled \tpre value for a given P/E-cycle count and retention age. 
At runtime, the SSD controller queries the RPT.

\begin{figure}[!h]
    \centering
    \includegraphics[width=\linewidth]{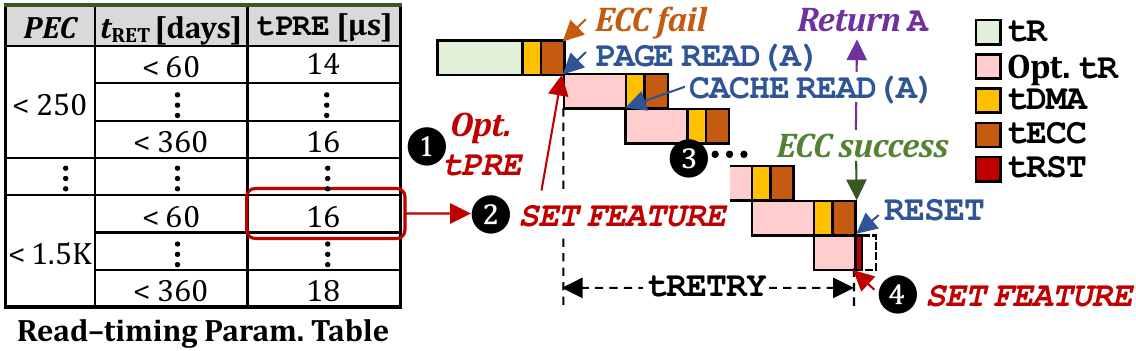}
    \vspace{-2em}
    \caption{Read-retry latency reduction in \arr.}
    \vspace{-1em}
    \label{fig:arr}
\end{figure}

\fig{\ref{fig:arr}} illustrates how the SSD controller can reduce the read-retry latency using \arr.\footnote{We assume that \prr is already implemented.}
Once a read failure occurs, \bcirc{1} \arr first decides the appropriate \tpre reduction amount by querying the RPT using the P/E-cycle count and retention age of the corresponding block.\footnote{A regular SSD already keeps track of the P/E-cycle count and retention age of each block (i.e., the information necessary to determine the best \tpre value) in order to ensure SSD lifetime and reliability (e.g., for wear leveling~\cite{chang-sac-2007}, periodic refresh of stored data~\cite{cai-inteltechj-2013, cai-iccd-2012, ha-ieeetcad-2015}, or optimal \vref{} prediction~\cite{cai-hpca-2015, luo-hpca-2018, luo-ieeejsac-2016, luo-sigmetrics-2018}).}
\bcirc{2} \arr then changes the target chip's \tpre value by issuing a \texttt{SET}~\texttt{FEATURE} command, and
\bcirc{3} repeats performing retry steps until the page is successfully read without any uncorrectable error or until all possible retry steps are exhausted.
Finally, \bcirc{4} \arr rolls back the target chip's \tpre value to the default \tpre value for future operations.

As explained above, in \arr, the SSD controller changes the \tpre value \emph{only once} (\bcirc{2}) for a read-retry operation, i.e., it performs \emph{all} the retry steps using the same \tpre value.
Doing so \emph{does not} increase the number of performed retry steps as long as the best \tpre value for a given ($PEC, t_\text{RET}$) is correctly found via offline profiling. 
As we observe in \sect{{\ref{subsec:margin}}}, the target page's RBER drastically decreases in the final retry step due to the use of near-optimal \vref{} values, while ECC decoding would fail anyway in all previous steps \emph{even if the default \tpre value were used}. 
In other words, with accurate profiling, the \tpre reduction does not affect the failure of the previous steps, while guaranteeing the success of the final retry step.

With \arr, \trr can be expressed as follows:
\begin{equation}
    \trr = \text{\texttt{tSET}} + \rho \times N_\text{RR} \times \tr + \tdma + \tecc
    \label{eq:trr_final}
\end{equation}
where \texttt{tSET} is the latency of the \texttt{SET}~\texttt{FEATURE} command for adjusting \tpre, and $\rho$ is reduction ratio in \tr ($0 < \rho \leq 1$).
As \tr is the dominant factor in \trr, \arr can provide considerable performance improvement.
For example, a 25\% \tr reduction ($=$22.5~\usec) under a 1-year retention age at 2K P/E cycles is easily possible (\sect{\ref{sssec:rrr}}).
Note that \texttt{tSET} is almost negligible (e.g., $<$ 1~\usec~\cite{micron-datasheet-2010}) compared to the total amount of \tr reduction. 

\head{Overhead}
\arr requires only small changes to the SSD controller to adjust \tpre based on the RPT.
The storage overhead of the RPT to store the best \tpre values for tens of (\pec, \tret) combinations is also very small.
For example, with 36 ($PEC, t_\text{RET}$) combinations, we estimate the table size to be only 144 bytes per chip.
Like other metadata related to each NAND flash chip (e.g., the P/E-cycle count of each block in the chip), the RPT can be stored in a specific page of each NAND flash chip and fetched into internal SRAM or DRAM at boot time so that the SSD controller can quickly access the RPT once a read-retry occurs.

\arr might potentially increase the number of retry steps for outlier pages that exhibit high RBER in the final retry step even with the default \tpre value.
In the worst case, i.e., when a read-retry fails\footnote{As explained in \sect{\ref{subsec:reliability_mgmt}}, a read-retry operation fails if the page's RBER is higher than the ECC capability even after trying all available sets of \vref{} values prescribed by manufacturers.} with reduced \tpre, \arr needs to perform a read-retry operation on the same page using the default \tpre value since the previous read-retry operation might not have failed with the default \tpre value.
However, the extremely-low probability of such a case (we never detect such a case in our characterization of more than $10^7$ pages) makes the potential performance overhead of \arr negligible.
Note that we also incorporate a safety margin for outlier pages into the reduced \tpre values, as shown in \fig{\ref{fig:best}}.

\begin{figure*}[!b]
    \centering
    \includegraphics[width=\linewidth]{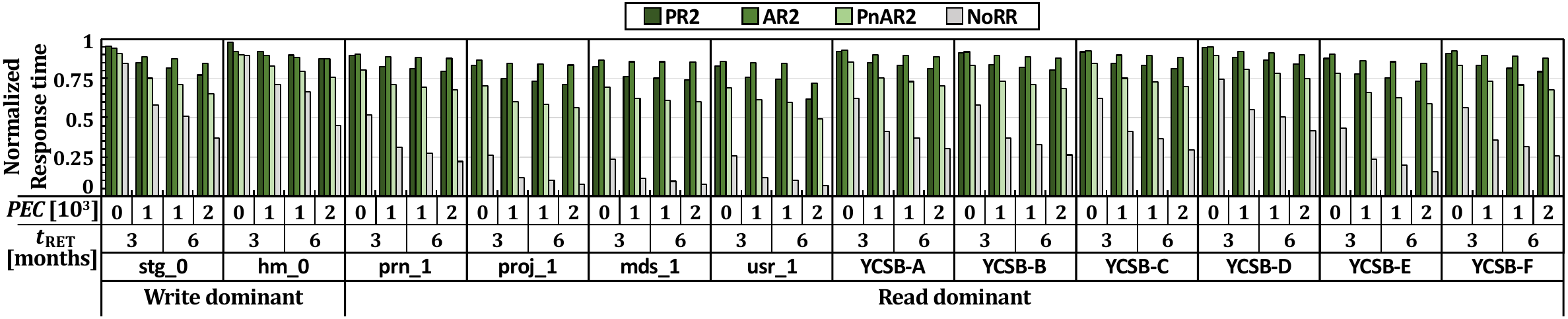}
    \vspace{-2em}
    \caption{Response-time (RT) reduction of our proposal under different \pec (unit: $\times10^3$) and \tret (unit: months).}
    \label{fig:eval}
\end{figure*}

\begin{figure*}[!b]
    \centering
    \includegraphics[width=\linewidth]{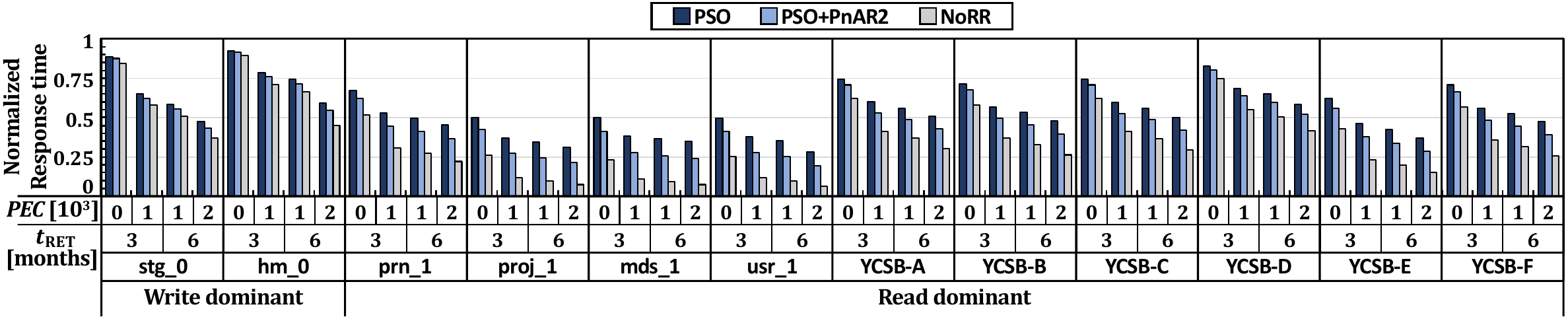}
    \vspace{-2em}
    \caption{Performance improvement of our proposal when combined with an existing read-retry mitigation scheme~\cite{shim-micro-2019}.}
    \label{fig:comp}
\end{figure*}

\vspace{-.1em}
\section{System-level Evaluation~\label{sec:eval}}
\vspace{-.1em}
We evaluate the impact of \prr and \arr on system performance using a state-of-the-art SSD simulator~\cite{tavakkol-fast-2018, mqsim-git} and twelve storage I/O traces from two representative benchmark suites. 

\vspace{-.3em}
\subsection{\textbf{Methodology}\label{subsec:method}}
\vspace{-.2em}
We evaluate the effectiveness of~\prr and \arr using MQSim~\cite{tavakkol-fast-2018, mqsim-git}, an open-source multi-queue SSD simulator. 
We faithfully extend MQSim based on our real-device characterization results to simulate more realistic read-retry characteristics of modern SSDs.
We modify the NAND flash model of MQSim such that each simulated block operates exactly the same as one of the real blocks that we test in \sect{\ref{sec:prof_results}}.
We randomly select real tested blocks and map each of them to a simulated block.
We modify the data structure of each simulated block to contain a lookup table for the number of read-retry steps at a certain P/E-cycle count and retention age, which we profile from the corresponding real block.
As MQSim maintains the P/E-cycle count and programming time of each page, a simulated block can accurately emulate the same read-retry behavior as the corresponding real block for every read.
We simulate a 512-GiB SSD that contains 4 channels, 4 dies per channel, and 2 planes per die. 
A plane consists of 1,888 blocks, each of which has 576 16-KiB pages.
We assume an ECC engine that corrects up to 72 bit errors per 1-KiB codeword within \tecc$=$20~\usec.
\tab{\ref{tab:config}} summarizes the timing parameters of our simulated NAND flash chip, which we obtain from the real NAND flash chips used in our characterization.
The I/O rate is set to 1Gb/s, i.e., \tdma = 16~\usec for a 16-KiB page.

\begin{table}[!h]
	\centering
	\vspace{-.5em}
	\caption{NAND flash timing parameters.}
	\vspace{-1em}
	\resizebox{\columnwidth}{!}{%
	\begin{tabular}{P{2cm};P{2cm}||P{2cm};P{2cm}}
		\toprule
		\textbf{Parameter} & \textbf{Time} & \textbf{Parameter} & \textbf{Time} \\
		\midrule
		\textbf{\tr (avg.)\footnotemark} & 90~\usec & \textbf{\tprog} & 700~\usec \\
		\textbf{\tpre} & 24~\usec & \textbf{\tbers} & 5~ms \\
		\textbf{\teval} & 5~\usec & \textbf{\texttt{tSET}} & 1~\usec \\
		\textbf{\tdisch} & 10~\usec & \textbf{\texttt{tRST}} & 5~\usec for read\\
  		\bottomrule
	\end{tabular}}
	\vspace{-.4em}
	\label{tab:config}%
\end{table}%

\begin{table}[!b]
	\centering
	\vspace{-0.6em}
	\caption{I/O characteristics of the evaluated workloads.}
	\vspace{-1em}
	\resizebox{1.0\columnwidth}{!}{%
	\begin{tabular}{P{1.4cm};P{1.2cm};P{1.2cm}||P{1.4cm};P{1.2cm};P{1.2cm}}
		\toprule
		\multirow{2}{*}{\textbf{Workload}} & \textbf{Read} & \textbf{Cold} & \multirow{2}{*}{\textbf{Workload}} & \textbf{Read} & \textbf{Cold} \\
		& \textbf{ratio} & \textbf{ratio} & & \textbf{ratio} & \textbf{ratio} \\
		\midrule
		\textbf{\stg} & 0.15 & 0.38 & \textbf{\yc{A}} & 0.98  &  0.72\\
		\textbf{\hm} & 0.36 & 0.22 & \textbf{\yc{B}} & 0.99 &  0.59 \\
		\textbf{\prn} & 0.75 & 0.72 & \textbf{\yc{C}} & 0.99  &  0.6 \\
		\textbf{\proj} & 0.89 & 0.96 & \textbf{\yc{D}} & 0.98 &  0.58 \\
		\textbf{\mds} & 0.92 & 0.98 & \textbf{\yc{E}} & 0.99 &  0.98 \\
		\textbf{\usr} & 0.96 & 0.73 & \textbf{\yc{F}} & 0.98 &  0.87 \\
		\bottomrule
	\end{tabular}}
	\vspace{-1.2em}
	\label{tab:workload}%
\end{table}%

\footnotetext{As explained in \sect{\ref{subsec:nand_opr}}, in multi-level cell NAND flash memory, \tr varies depending on $N_\text{SENSE}$, the number of sensing times required for reading a page (Equation~\eqref{eq:tr}).
In TLC NAND flash memory, $N_\text{SENSE}=\langle 2, 3, 2\rangle$ for $\langle$LSB (least-significant bit), CSB (central-significant bit), MSB (most-significant bit)$\rangle$ pages~\cite{cai-procieee-2017, cai-insidessd-2018}.}

We evaluate twelve workloads from two benchmark suites, Microsoft Research Cambridge (MSRC) Traces~\cite{narayanan-fast-2008} and Yahoo! Cloud Service Benchmark (YCSB)~\cite{cooper-socc-2010}, which are widely used for performance evaluation of storage systems~\cite{cai-iccd-2012, cai-sigmetrics-2014, luo-hpca-2018, luo-sigmetrics-2018, luo-msst-2015, kim-asplos-2020, liu-asplos-2019, shim-micro-2019, tavakkol-fast-2018, tavakkol-isca-2018}.
MSRC Traces consist of 36 block I/O traces that are collected from enterprise servers during one week. 
We select six traces from the 36 total traces to have different I/O characteristics in terms of \emph{read ratio} and \emph{cold ratio}, as summarized in \tab{\ref{tab:workload}}.
Read ratio is the fraction of read requests in a workload.
Cold ratio is the fraction of read requests whose target page is never updated during the entire execution of the workload. Such a page is programmed only once and thus experiences a long retention age compared to frequently-updated pages (i.e., write-hot pages).

\subsection{SSD Response Time \label{subsec:eval_results}}
We compare five SSD configurations with different read-retry mitigation schemes: 1) \base, 2) \prssd, 3) \arssd, 4) \parssd, and 5) \norr.
\base is a high-end SSD that 1) adopts out-of-order I/O scheduling~\cite{jung-hpca-2014, tavakkol-isca-2018} and program/erase suspension~\cite{kim-atc-2019, wu-fast-2012} techniques to provide high read performance, and 2) performs regular read-retry operations (described in \fig{\ref{fig:prr}(a)}).
\prssd and \arssd are SSDs that implement each of our two techniques alone on top of \base, respectively.
\parssd (\emph{Pipelined and Adaptive Read-Retry}) is an SSD that integrates both \prr and \arr to minimize the read-retry latency.
\norr is an \emph{ideal} SSD where no read-retry occurs, showing the upper bound of eliminating read-retry on SSD performance.
\fig{\ref{fig:eval}} compares the performance (average response time) of all five SSD configurations, normalized to \base, under different operating conditions. 

We make five key observations from \fig{\ref{fig:eval}}.
First, our two techniques, either alone or when combined, significantly improve the I/O performance of a modern SSD.
\prr and \arr reduce SSD response time by up to 38.3\% and 18.1\% (17.7\% and 11.9\% on average across all workloads) compared to \base, respectively.
\parssd provides higher improvements by using both \prr and \arr, enabling up to 51.8\% (28.9\% on average) SSD response time reduction over \base.
Second, \prr and \arr improve SSD performance in a \emph{synergistic manner}.
The average SSD response time improvement of \parssd (28.9\% on average across all workloads) is higher than simple aggregation of the individual gains of \prssd and \arssd.
Third, the worse the operating conditions, the larger the performance gain of the proposed techniques.
For example, under a 6-month retention age at 2K P/E cycles, \parssd reduces the average SSD response time across all workloads by 35.2\% over \base.
Fourth, our proposal is highly effective under read-dominant workloads, but it also provides considerable performance improvement for workloads with many writes.
For \textsf{stg\_0}, whose read ratio is only 0.15, \parssd improves SSD performance by 34\% under a 6-month retention age at 2K P/E cycles and by 18.7\% on average under all operating conditions.
The results suggest that, although the page-read latency (\tr) is much shorter than the page-program latency (\tprog) and block-erasure latency (\tbers), it is important to optimize read-retry as frequent read-retry operations with multiple retry steps can cause read requests to bottleneck SSD performance.
Fifth, \prr and \arr are effective at mitigating the performance overhead of read-retry, but there still exists large room for improvement. 
\parssd reduces the response-time gap between \base and \norr by 41\% on average across all workloads, but still exhibits a 2.37$\times$ higher average response time compared to the ideal \norr.

Based on our observations, we conclude that our proposal is effective at improving SSD performance by mitigating the significant overheads of read-retry. 
We also find that there is still large room for improvement to optimize \parssd, which motivates us to combine \prr and \arr with existing read-retry optimization techniques that aim to reduce the \emph{number} of read-retry operations.

\vspace{-1em}
\subsection{Comparison to Prior Work\label{subsec:comp}}
To evaluate the effectiveness of our proposal when combined with existing read-retry optimization techniques, we compare two SSD configurations that adopt a state-of-the-art read-retry mitigation technique~\cite{shim-micro-2019}, called PSO (Process Similarity-aware Optimization). 
PSO reduces the number of retry steps by reusing \vref{} values that are recently used for a read-retry operation on other pages exhibiting similar error characteristics with the target page to read.
We simulate two PSO-enabled SSDs using MQSim: 1) \textsf{PSO}, which adopts only the PSO technique over \base, and 2) \textsf{PSO+PnAR2}, which integrates \prr and \arr on top of \textsf{PSO}.  
\fig{\ref{fig:comp}} compares the average response time of the two SSDs under different conditions. 
All values are normalized to \base.

We make three major observations.
First, although \textsf{PSO} significantly reduces the average response time over \base, its performance is still far from the ideal \norr.
In particular, the average response time of \textsf{PSO} is up to 4.31$\times$ (1.92$\times$ on average) that of \norr in read-dominant workloads.
Second, \prr and \arr further improve SSD performance significantly when implemented on top of the PSO technique.
Compared to \textsf{PSO}, \textsf{PSO+PnAR2} reduces the average response time by up to 31.5\% (17\% on average) in read-dominant workloads and by up to 9.4\% (3.6\% on average) in write-dominant workloads.
Third, \textsf{PSO+PnAR2}'s average response time is 1.6$\times$ that of the ideal \nrr in read-dominant workloads. 
This shows that there is still some more room for optimizing read-retry in future work.

We conclude that our proposal effectively complements existing techniques to minimize the read-retry overhead and thus significantly improves the performance of modern SSDs.
We believe that \prr and \arr are quite promising as their large performance benefits come with almost negligible overheads.

\vspace{-.5em}
\section{Discussion\label{sec:discussion}}
We briefly discuss the potential impact of our proposals on future research to optimize SSD read performance.

\head{Latency Reduction for Regular Reads}
A modern SSD's high ECC capability enables \arr to use a page's ECC-capability margin for reducing the sensing latency.
Although \arr is used for only reducing the latency of a read-retry, we expect that its key idea can be used also for reducing the latency of a regular page read (i.e., a page read requiring no read-retry). For example, if we can accurately estimate a page's RBER and near-optimal \vref{} values using an accurate error model (which could be obtained via extensive real-device characterization as done in~\cite{cai-date-2012, cai-date-2013, cai-hpca-2015, cai-iccd-2013, cai-inteltechj-2013, cai-dsn-2015, cai-iccd-2012, cai-procieee-2017, cai-insidessd-2018, cai-sigmetrics-2014, luo-hpca-2018, luo-ieeejsac-2016, luo-sigmetrics-2018, shim-micro-2019}), it would enable us to safely reduce read-timing parameters for regular reads. 

\head{Further Reduction of Read-Retry Latency} 
We expect that the key idea of \prr, i.e., speculatively starting a new retry step (assuming that the current retry step is likely to fail), can be extended to further reduce the read-retry latency.
For example, if a page to read is likely to exhibit high RBER that would exceed the ECC capability, the SSD controller can speculatively start read-retry steps without reading the page using default timing parameters (which would likely fail).
By using an accurate error model that can predict whether or not a page read would fail (e.g., as in~\cite{cai-date-2012, cai-date-2013, cai-hpca-2015, cai-iccd-2013, cai-inteltechj-2013, cai-procieee-2017, cai-insidessd-2018, luo-hpca-2018, luo-ieeejsac-2016, luo-sigmetrics-2018, shim-micro-2019}), such an approach could further reduce the effective read-retry latency without penalty.

\vspace{-.5em}
\section{Related Work}
To our knowledge, this paper is the first to 1) provide a detailed and rigorous understanding of the read-retry behavior and the effect of reducing read-timing parameters in modern NAND flash memory by characterizing a large number of real 3D NAND flash chips, and 2) propose two new techniques that effectively reduce the read-retry latency by exploiting advanced features of modern SSDs.
We briefly discuss closely related prior work that aims to mitigate the read-retry overhead and improve system performance by exploiting the reliability margin in memory devices, specifically NAND flash memory and DRAM.

\head{Read-Retry Mitigation} 
Prior works~\cite{cai-inteltechj-2013, cai-iccd-2012, ha-ieeetcad-2015} propose to refresh a page (i.e., reset the page's retention age) before the page's RBER increases beyond the ECC capability to avoid a read failure that, in turn, causes a read-retry.
Refresh-based approaches might be able to reduce the number of read-retry operations, but can negatively affect the overall SSD performance by wasting bandwidth and/or increasing wear due to refresh operations.
As explained in \sect{\ref{subsec:reliability_prob}}, a page's RBER rapidly increases beyond the ECC capability in modern 3D NAND flash memory, which incurs a read-retry operation even under a zero retention age as shown in \fig{\ref{fig:rr_dist}}. 
To avoid read-retry, therefore, refresh-based approaches should refresh (i.e., read and rewrite) soon-to-be-read pages \emph{very frequently}, which could introduce significant performance overheads and wear, as shown in~\cite{cai-inteltechj-2013, cai-iccd-2012}.

Several prior works~\cite{cai-hpca-2015, cai-iccd-2013, luo-ieeejsac-2016, luo-hpca-2018, luo-sigmetrics-2018, nie-dac-2020, shim-micro-2019} propose to keep track of pre-optimized \vref{} values for each block to use the \vref{} values for future read requests.
By starting a read (and retry) operation with pre-optimized \vref{} values close to the optimal read-reference voltage \vopt, they significantly reduce the number of read-retry steps.
Unfortunately, read-retry is a fundamental problem that is hard to completely eliminate in modern SSDs.
While the existing techniques reduce the \emph{number} of read-retry steps, our techniques effectively reduce \emph{the latency for performing the same number of retry steps}, which makes our techniques complementary to these existing techniques, as shown in \sect{\ref{subsec:comp}}. 
Considering the low overhead of our techniques, they can be easily combined with existing techniques to minimize the read-retry overhead that is expected to become larger in emerging NAND flash memory (e.g., 3D QLC NAND flash memory).

A concurrent study~\cite{li-micro-2020} with ours proposes a new \vopt prediction technique to reduce the number of retry steps.
The key idea is to store predefined bit patterns in a set of spare cells in each page, called \emph{Sentinel} cells, so as to accurately estimate the page's current error characteristics (and \vopt) based on errors incurred in the predefined bit patterns. Doing so allows the SSD controller to try near-optimal \vref{} values right after the first regular page read, which significantly reduces the average number of read-retry steps (from 6.6 to 1.2).
Both of our proposed techniques, \prr and \arr, can complement the Sentinel-based approach~\cite{li-micro-2020} as well as other read-retry mitigation techniques~\cite{cai-hpca-2015, cai-iccd-2013, luo-ieeejsac-2016, luo-hpca-2018, luo-sigmetrics-2018, nie-dac-2020, shim-micro-2019} that aim to reduce the \emph{number} of retry steps.
First, once the SSD firmware accurately identifies the optimal \vref{} values using the Sentinel cells, it is possible to reduce the latency of the following retry step(s) using \arr, i.e., applying reduced read-timing parameters.
Second, \prr reduces the latency of a read-retry operation when the Sentinel-based \vopt prediction fails (as shown in \cite{li-micro-2020}, the Sentinel-based approach cannot completely avoid multiple retry steps), by speculatively issuing a new retry step.

\head{Full Utilization of Reliability Margin in Memory Devices}
There is a large body of prior work to improve DRAM performance and energy consumption by leveraging variation in access latency~\cite{chang-sigmetrics-2016, kim-iccd-2018, lee-sigmetrics-2017, lee-hpca-2015, wang-micro-2018, hassan-hpca-2017, koppula-micro-2019, kim-hpca-2018, kim-hpca-2019, chang-sigmetrics-2017, luo-isca-2020, hassan-hpca-2016, zhang-hpca-2016, chandrasekar-date-2014} and data retention times~\cite{liu-isca-2012, qureshi-dsn-2015, das-dac-2018, khan-micro-2017, khan-cal-2016, khan-dsn-2016, khan-sigmetrics-2014, liu-ISCA-2013, venkatesan-HPCA-2006, patel-isca-2017} across DRAM cells.
These works clearly show that manufacturers set timing parameters of a device conservatively to ensure correct operation of outlier cells, even though a significant majority of cells can reliably operate with lower timing parameters.
Our work shows that there also exists a large reliability margin in modern NAND flash memory and demonstrates that the large margin can be used for reducing the read latency through careful real-device characterization.

A few prior works~\cite{liu-fast-2012, pan-hpca-2012, shim-micro-2019} propose to improve the write performance of SSDs by fully exploiting the underutilized ECC capability.
For example, Liu et al.~\cite{liu-fast-2012} propose to program a soon-to-be-updated page with coarse-grained voltage control, which greatly reduces the program latency.
As the page will not experience a long retention age, programming the page with fine-grained voltage control unnecessarily increases the program latency while leaving a large ECC-capability margin for the page.
Similarly, Shim et al.~\cite{shim-micro-2019} propose to reduce the number of verify steps in a page-program operation, if the target WL is more robust to error than other WLs.
By leveraging processing variation in modern NAND flash memory, they safely reduce the program latency for a large number of WLs.
None of these techniques, however, leverage the ECC-capability margin to improve SSD read latency, which is often more critical for many key applications in modern computing systems.
\section{Conclusion}
This paper proposes two new read-retry mechanisms, \prr and \arr, which significantly improve SSD performance by reducing the latency of read-retry operations.
We identify new opportunities to optimize the read-retry latency by exploiting advanced architectural features widely adopted in modern SSDs.
Through extensive real-device characterization of modern 3D TLC NAND flash chips, we demonstrate that it is possible to use the large ECC-capability margin present in the final retry step for reducing the latency of each retry step.
Our results show that our techniques effectively improve SSD performance and complement existing techniques.  
We hope that our new findings on the read-retry behavior and the reliability impact of reducing read-timing parameters in modern NAND flash-based SSDs inspire new mechanisms to further improve SSD latency and performance, which are critical for modern data-intensive workloads.

%Acknowledgement
\begin{acks}
We would like to thank our shepherd Dan Tsafrir and anonymous reviewers for their feedback and comments.
We thank the SAFARI Research Group members for feedback and the stimulating intellectual environment they provide.
We thank our industrial partners, especially Google, Huawei, Intel, Microsoft, and VMware, for their generous donations. 
This work was in part supported by Samsung Research Funding \& Incubation Center of Samsung Electronics, Republic of Korea under Project Number SRFC-IT2002-06, and by the National Research Foundation of Korea (NRF) grant funded by the Ministry of Science and ICT (MSIT) (NRF-2020R1A6A3A03040573).
The ICT at Seoul National University provided research facilities for this study.
(\emph{Co-corresponding Authors: Jisung Park, Jihong Kim, and Onur Mutlu})
\end{acks}

%References
%Balance the last page
\balance
\bibliographystyle{ACM-Reference-Format-num}
\bibliography{reference}

\end{document}